\documentclass[sigconf]{acmart}
\usepackage[utf8]{inputenc}
\usepackage{textgreek}
\usepackage{xcolor}
\usepackage{amsmath}
\usepackage[ruled, vlined, linesnumbered]{algorithm2e}
\usepackage{bm}
\usepackage{enumitem}
\usepackage{lipsum} 
\usepackage[most]{tcolorbox}
\usepackage{tikz}
\usepackage{pgfplots}
\pgfplotsset{compat=1.18}
\usetikzlibrary{arrows.meta, positioning, backgrounds, fit}
\usepackage[aboveskip=-4pt]{subcaption}
\usepgfplotslibrary{fillbetween}
\usepackage{amsthm}
\usepackage{mdframed}
\usepackage{seqsplit}
\usepackage{placeins}
\usepackage{multirow}
\usepackage{colortbl}

\usepackage{booktabs}
\usepackage{array}
\usepackage{adjustbox}
\usepackage{tabularx}
\usepackage{array}
\newcolumntype{R}{>{\raggedleft\arraybackslash}X} 

\theoremstyle{definition}

\newcommand{\eat}[1]{} 

\newcommand{\stitle}[1]{\vspace*{0.2em}\noindent{\bf #1.\/}}

\definecolor{hlbg}{RGB}{200,240,240}

\newcommand{\ourbench}{\textsc{VoteGEO-Bench}\xspace}
\newcommand{\ussimple}{\textit{us-election-24}\xspace}
\newcommand{\usrich}{\textit{ai-persuasion}\xspace}
\newcommand{\jp}{\textit{jp-election-26}\xspace}

\newcommand{\google}{\texttt{Google-SERP}\xspace}
\newcommand{\gptfivemini}{\texttt{\seqsplit{gpt-5-mini}}\xspace}
\newcommand{\gptfourmini}{\texttt{\seqsplit{gpt-4o-mini}}\xspace}
\newcommand{\grok}{\texttt{\seqsplit{grok-4-1-fast}}\xspace}
\newcommand{\grokx}{\texttt{\seqsplit{grok-with-x}}\xspace}
\newcommand{\gemini}{\texttt{\seqsplit{gemini-2.5-flash}}\xspace}

\newcommand{\cD}{\mathcal{D}}

\newenvironment{customlegend}[1][]{%
    \begingroup
    \csname pgfplots@init@cleared@structures\endcsname
    \pgfplotsset{#1}%
}{%
    \csname pgfplots@createlegend\endcsname
    \endgroup
}%

\def\addlegendimage{\csname pgfplots@addlegendimage\endcsname}

\makeatletter
\newcommand\footnoteref[1]{\protected@xdef\@thefnmark{\ref{#1}}\@footnotemark}
\makeatother

\definecolor{RedPPAlt}{HTML}{DC143C}
\definecolor{Red}{HTML}{E81123}
\definecolor{Orange}{HTML}{FFB000}
\definecolor{Green}{HTML}{009E49}
\definecolor{LightBlue}{HTML}{00BCF2}
\definecolor{DeepBlue}{HTML}{001BA3}
\definecolor{Pink}{HTML}{7C3AED}
\definecolor{LightGreen}{HTML}{8BC63E}
\definecolor{Gray}{HTML}{6B7280}

\newcounter{myremark}
\renewcommand{\themyremark}{\arabic{myremark}}

\newtcolorbox{remark}[1][]{
    enhanced,
    breakable,
    before skip=10pt,
    after skip=10pt,
    colback=gray!5,
    colframe=navyblue!70!black,
    boxrule=0.3pt,
    leftrule=3pt,
    left=8pt,
    right=5pt,
    top=5pt,
    bottom=5pt,
    sharp corners,
    fonttitle=\bfseries,
    coltitle=black,
    title={Remark \refstepcounter{myremark}\themyremark #1},
    attach title to upper,
    after title={.\hspace{0.5em}},
}
\definecolor{navyblue}{RGB}{0, 0, 128}
\newcommand{\trref}[1]{Appendix~\ref{appendix}}

\AtBeginDocument{%
  }

\begin{document}

\title{\textsc{VoteGEO-Bench}: Measuring the Impact of Generative Engine Optimization on Electoral Visibility}

\author{Shiqi Zhang}
\orcid{0000-0002-7155-9579}
\affiliation{%
  \institution{National University of Singapore}
  \country{Singapore}
}
\email{sqzhang@nus.edu.sg}

\author{Xiaokui Xiao}
\orcid{0000-0003-0914-4580}
\affiliation{%
  \institution{National University of Singapore}
    \country{Singapore}
}
\email{xkxiao@nus.edu.sg}

\begin{abstract}
Large language models (LLMs) are increasingly used to obtain election information, and the web sources they use can shape the responses presented to voters. Existing work on LLMs in political settings mainly studies political bias in model outputs and AI-driven political persuasion, while much less is known about how external web content affects the political information presented by LLMs. Generative engine optimization (GEO) focuses on rewriting web content to improve its visibility in generated responses. Existing GEO research typically takes a fixed webpage pool as input and studies how rewriting changes page visibility in generated responses, providing a natural framework for studying this question.
To investigate the potential risks of GEO in voting scenarios (\textsc{VoteGEO}), we introduce \ourbench, a benchmark built on the 2024 U.S. presidential election and the 2026 Japanese House of Representatives election, combining election-related queries, historical webpages, and voter personas.

We first analyze the sources cited by generative engines, finding that many citations come from lower-cost owned and open-platform sources and that some webpages often remain visible across days. We then benchmark 11 rewriting strategies across 4 categories and study whether specifying a target candidate or party in the rewrite matters, which strategies work across generative engines and elections, and how voter personas affect exposure. Target-unspecified rewriting rarely increases target exposure, whereas target-specified rewriting more often does so. Notably, strategies that improve page organization or strengthen a position through argumentation show stronger gains than those that add supporting evidence or mainly change wording. For the two strongest strategies, rewriting gains remain broadly positive across the evaluated U.S. personas, while persona conditioning shows no consistent tendency to increase or decrease these gains. In contrast, explicit partisan identities can create baseline exposure gaps larger than the gain from rewriting a single webpage. When partisan information is only implicit in the voter profile, baseline exposure shifts vary with how strongly the engine responds to the implicit partisan signal.
\end{abstract}

\maketitle
\section{Introduction}

Large language models (LLMs) are increasingly used to obtain election-related information, including candidate comparisons and voting advice~\cite{medina2026who,ap2025ranspecial}. Unlike traditional search engines that rank and present webpages, generative engines such as ChatGPT and Gemini retrieve {web sources} and use their content to generate responses. Existing work on LLMs in political settings mainly follows two lines. One studies political bias and alignment in LLM responses across models and political contexts~\cite{batzner2025germanpartiesqa,kim2026polar,damiao2026evidence,barmettler2026invisible,cen2025large}. The other studies whether LLM-generated political messages can influence voters and which factors drive political persuasion~\cite{lin2025persuading,hackenburg2025levers}. These studies focus on LLM outputs and their effects on users. Much less is known about \textit{how external web content shapes political information in generative engine responses}.

\textit{Generative engine optimization} (GEO) rewrites web content to improve its visibility in generative engine responses~\cite{geo,c-seo}. Existing GEO research typically takes a fixed webpage pool as input and studies how rewriting changes page visibility in generated responses~\cite{geo,c-seo,ragdoll,tang2025stealthrank,autogeo,raid_white}, providing a natural framework for studying this question. 
To investigate the potential risks of GEO in voting scenarios, which we call \textsc{VoteGEO}, it remains two sets of questions. First, which source channels do generative engines use, how accessible are they to content publishers, and how long do the same webpages remain visible? Second, how does webpage rewriting affect candidate or party exposure in generated responses? Specifically, does specifying a target candidate or party in the rewrite matter, which strategies work across generative engines and elections, and how do voter personas affect baseline exposure and rewriting gains?
To answer these questions, we introduce \ourbench, a benchmark built on the 2024 U.S. presidential election~\cite{cen2025large} and the 2026 Japanese House of Representatives election~\cite{miyazaki2026ai}. After preprocessing, \ourbench contains 28,200 election-related query cells across days and engines and 17,251 webpages. We further incorporate 3 categories of voter personas, including demographic identities, partisan identities, and questionnaire-derived profiles constructed from pre-treatment responses of 2,306 U.S. participants~\cite{lin2025persuading}.

We first analyze which sources generative engines use and how long the same webpages remain visible. We find that many cited sources come from lower-cost channels, including candidate or party websites and open platforms such as Reddit. For queries about relatively stable topics, some webpages repeatedly appear across days, whereas those for current events turn over much faster. Together, these findings highlight potential risks through lower-cost source channels, with longer exposure windows for stable queries and faster source turnover for current events.
We then benchmark 11 rewriting strategies across 4 categories. Structure strategies improve page organization or reformat webpages into more LLM-friendly layouts. Evidence strategies add quotations, statistics, or source attributions to support existing claims. Style strategies modify wording without adding substantive content. These three categories follow existing GEO work~\cite{geo,c-seo}. Motivated by political persuasion research~\cite{lin2025persuading,hackenburg2025levers}, we further introduce 2 Argumentation strategies, which strengthen a position by emphasizing its distinctions from plausible alternatives or explicitly addressing opposing arguments.

Our results reveal three main patterns. First, target-unspecified rewriting rarely increases target exposure, whereas target-specified rewriting more often does so. This distinction is important because election webpages often discuss multiple parties. In the U.S. setting, 92\% of retrieved webpages mention both major parties, so rewriting without an explicit target does not indicate which party's exposure should increase. Second, Structure and Argumentation strategies show stronger gains than Evidence and Style strategies across generative engines and elections. Consistent with this pattern, one strategy from each of these two categories shows the strongest overall gains across settings and exposure metrics. Third, for these two strongest strategies, rewriting gains remain broadly positive across the 15 evaluated U.S. personas, while persona conditioning shows no consistent tendency to increase or decrease these gains. In contrast, explicit partisan identities can create baseline exposure gaps of up to 11.6 percentage points, larger than the gain from rewriting a single webpage. We further consider whether similar shifts appear when partisanship is only implicit in the voter profile. In this case, baseline exposure shifts vary with how strongly each engine responds to the implicit partisan signal.

To summarize, we make the following contributions:
\begin{itemize}[topsep=2pt,itemsep=1pt,parsep=0pt,partopsep=0pt,leftmargin=11pt]
    \item We introduce \ourbench, built on two elections with election queries, historical webpages, and voter personas (Section~\ref{sec:data}).

    \item We provide motivating insights into the potential risks of \textsc{VoteGEO} by analyzing source types and webpage visibility over time (Section~\ref{sec:motivating}).

    \item We organize 11 rewriting strategies into 4 categories, including 2 new Argumentation strategies (Section~\ref{sec:strategies}).

    \item We evaluate how target specification, rewriting strategies, and voter personas affect target exposure across engines and elections (Section~\ref{sec:exp}).
\end{itemize}
\section{Preliminaries}
\label{sec:prelim}

\subsection{\textsc{VoteGEO}}
\label{sec:formulation}

Generative engines first retrieve a set of webpages relevant to the query, often through web search, and then use these webpages as context to generate a response. Generative engine optimization (GEO) aims to improve the visibility of web content in such responses through content rewriting~\cite{geo,c-seo}. Existing GEO research typically separates retrieval from generation and focuses on the generation stage by fixing the retrieved webpage pool and studying how rewriting one webpage changes its visibility in the generated response~\cite{geo,c-seo,ragdoll,tang2025stealthrank,autogeo,raid_white}. We follow this problem setting and define a \textit{query cell} as an election-related query issued to a particular engine on a particular day. Each query cell $i$ has a query $q_i$ and a corresponding webpage pool $\cD_i$. Let $d_i \in \cD_i$ denote the webpage to be rewritten. The other webpages in $\cD_i$ remain fixed, and GEO studies whether rewriting $d_i$ improves its visibility in the generated response.

Existing GEO settings typically consider webpages centered on a single target entity, such as a product description page or a brand landing page~\cite{geo,c-seo}. In voting scenarios, however, a webpage may discuss multiple candidates or parties, so selecting a webpage does not uniquely determine whose exposure the rewrite should increase. \textsc{VoteGEO} therefore introduces a \textit{target} $t_i$, defined as a candidate or party mentioned in $d_i$, and studies whether rewriting $d_i$ can increase the exposure of $t_i$ in the generated response.

Given an engine $G$, let $y_i = G(q_i,\cD_i)$ denote the response before rewriting. Under \textit{target-specified} rewriting, the rewriting model is given both $d_i$ and $t_i$, whereas under \textit{target-unspecified} rewriting, it is given only $d_i$. Let $\bar{d}_i$ denote the rewritten webpage, and let $\bar{\cD}_i = (\cD_i \setminus \{d_i\}) \cup \{\bar{d}_i\}$ denote the webpage pool after replacement. The engine then produces $\bar{y}_i = G(q_i,\bar{\cD}_i)$. Let $\phi(t,y)$ denote an exposure measure for target $t$ in response $y$. The exposure gain for query cell $i$ is
\begin{equation}\label{eq:obj}
\Delta_i
=
\phi(t_i,\bar{y}_i)-\phi(t_i,y_i).
\end{equation}

Prior work on the mere exposure and illusory truth effects shows that repeated exposure can increase the perceived accuracy of information, even when the information is false~\citep{pennycook2018prior}. This motivates our focus on target exposure, which we measure textually without distinguishing favorable from unfavorable mentions.

\subsection{Related Work}
\label{sec:related}

\stitle{Political bias in LLMs}
Existing work studies political bias in LLM outputs across models and political contexts. Studies report a left-progressive lean in U.S. political content but more mixed patterns in South Korea~\citep{kim2026polar}, stronger alignment of GPT-4o with left-leaning German parties~\citep{batzner2025germanpartiesqa}, and overrepresentation of radical-right actors in European election queries~\citep{damiao2026evidence}. Other work finds that such patterns do not always persist for concrete voting decisions, such as Swiss referenda~\citep{barmettler2026invisible}. These studies examine political bias in model outputs, whereas we study how modifying external web content changes candidate or party exposure in generated responses.

\stitle{Voter persuasion with LLMs}
Prior work identifies several factors relevant to political persuasion with LLMs, including factual and evidence-rich information, personalization, and refutational counterargumentation. Factual and evidence-rich information repeatedly emerges as an important driver of persuasion~\citep{lin2025persuading,hackenburg2025levers,bai2025llmgenerated}, while the additional benefit of personalization is mixed across studies~\citep{matz2024potential,salvi2025conversational,lin2025persuading,hackenburg2025levers}. Refutational messages can also improve persuasion or reduce issue polarization~\citep{okeefe1999good,hruschka2026reducing}. These studies focus on how LLM responses influence voters. In contrast, we study how external content influences the information produced by LLMs.

\stitle{Generative engine optimization}
Existing GEO studies develop strategies for improving source visibility or item ranking in generative engine responses. Adversarial approaches use model-directed instructions or optimized content to manipulate preferences and rankings~\citep{ragdoll,black-hat-geo,tang2025stealthrank,DBLP:conf/emnlp/QianZJJW25,jin2026controlling}. Content optimization approaches instead rewrite wording or structure, learn generative engine preferences, or account for retrieval while preserving semantic fidelity~\citep{bardas2025white,geo,c-seo,raid_white,autogeo,ho2025rewrite}. Unlike prior GEO work that mainly optimizes source visibility or item ranking, we study target exposure within webpages that may discuss multiple candidates or parties.

\section{Data Collection and Preprocessing}
\label{sec:data}

We construct three datasets, \ussimple, \usrich, and \jp, from two electoral settings, the 2024 U.S. presidential election between the Democratic and Republican parties and the proportional-representation component of the 2026 Japanese House of Representatives election involving 11 parties. \ussimple and \jp contain election-related query cells and their corresponding webpage pools, while \usrich provides questionnaire information for constructing richer voter personas in the U.S. setting.  Table~\ref{tab:dataset_summary} summarizes the resulting query cells and webpage datasets. Additional details on query composition, persona specifications, and webpage preprocessing are provided in \trref{}.

\subsection{Data Sources}
\label{sec:data-source}
\ussimple is derived from the longitudinal corpus for the 2024 U.S. presidential election~\cite{cen2025large}. The original dataset contains 573 election-related queries collected on a near-daily basis from July 23 to November 12, 2024, together with historical Google Search results. 
\usrich is based on the voter-persuasion dataset~\cite{lin2025persuading}. We use the pre-treatment questionnaire responses of 2,306 U.S. participants in the 2024 presidential election, covering demographics, political engagement, voting criteria, and other voter characteristics.
\jp is derived from the longitudinal factorial dataset of \citet{miyazaki2026ai} for the 2026 Japanese House of Representatives election. The dataset contains 25 voting-advice queries, 44 demographic personas, and responses from five web-enabled LLMs collected daily from February 2 to February 8, 2026.

\subsection{Query Cells and Voter Personas}
\label{sec:data-query-persona}

\stitle{Query cells}
For \ussimple, we retain 260 of the 573 queries that are relevant to voter decision making, covering candidate evaluation and comparison, policy positions, election predictions, and related news. One example is {``Which US presidential candidate do voters believe will handle the economy the best?''} We use the 106 days from July 23 through election day on November 5, 2024, with Google Search as the engine. The first collection day contains SERP results for only 250 queries, resulting in 27,550 query cells.
For \jp, we retain all 25 voting-advice queries, consisting of 12 policy issues with two opposing stances for each issue and one control query without a policy stance. Each asks which political party the voter should vote for, optionally accompanied by a policy stance. For example, one query asks {``I think Japan should raise defense spending to at least 2\% of GDP. Which political party should I vote for?''} We retain four web-enabled LLMs as engines, \gptfourmini, \gptfivemini, \grok, and \grokx. We exclude \gemini because only 10 of its 7,700 responses contain parseable citation URLs. Two \gptfourmini collection days are unavailable in the source data, leaving 26 engine and day combinations and 650 query cells.

\stitle{Voter personas}
Personas form an independent conditioning dimension and can be combined with the same query cells. We consider three categories of voter personas: demographic identities, partisan identities, and questionnaire-derived profiles.
\ussimple provides 11 identity personas~\citep{cen2025large}, which we group into 9 demographic identities, covering 6 racial or ethnic identities and 3 gender identities, and 2 partisan identities, Democrat and Republican. We additionally include an Independent persona representing voters who identify with neither major party as a study-specific partisan identity. \usrich provides pre-treatment questionnaire responses from 2,306 U.S. participants, from which we use demographics, political engagement, and voting criteria to identify three interpretable subgroups and construct questionnaire-derived profiles~\citep{lin2025persuading}. \jp provides 44 demographic identities formed by crossing 2 genders, 11 regions, and 2 area types~\citep{miyazaki2026ai}.

\subsection{Webpage Pools}
\label{sec:data-webpages}

Following the fixed-pool GEO setting in Section~\ref{sec:formulation}, we construct a webpage pool $\cD_i$ for each query cell. Voting-related web content can change over time during an election, so our goal is to reconstruct a query-relevant snapshot of the web content available around the collection day. To this end, we use the historical search results and citations available in the source datasets to collect relevant URLs for each query cell.

For \ussimple, each pool consists of the persona-free historical \google result URLs for the corresponding query and day. For \jp, each user prompt combines a query with a persona and, where applicable, a policy stance. To construct a comprehensive query-relevant webpage pool while mitigating dependence on any single persona or stance, we aggregate citation URLs for the same engine, query, and day across all 44 personas and the two opposing stances of each policy issue.
For each URL associated with a query cell, we collect the webpage content from as close as possible to its collection day. We first use the closest available Wayback Machine snapshot, followed by live-page and headless-browser fallbacks when no archived snapshot is available. We then extract the main textual content from each webpage.

The resulting pools contain 9.15 usable webpages per query cell on average for \ussimple and 63.1 for \jp. Across all query cells, the final datasets contain 11,501 and 5,750 distinct webpages related to the U.S. and Japanese elections, respectively.


\begin{table}[t]
\centering
\small
\caption{Summary of the two election datasets.}
\label{tab:dataset_summary}
\vspace{-2mm}
\setlength{\tabcolsep}{3.5pt}
\renewcommand{\arraystretch}{1.12}

\begin{adjustbox}{max width=\columnwidth}
\begin{tabular}{lrrrrrr}
\toprule
\textbf{Dataset}
& \multicolumn{4}{c}{\textbf{Query cells}}
& \multicolumn{2}{c}{\textbf{Webpages}} \\
\cmidrule(lr){2-5}
\cmidrule(lr){6-7}

&
\textbf{Query}
&
\textbf{Day}
&
\textbf{Engine}
&
\textbf{Total}
&
\textbf{Avg./cell}
&
\textbf{Distinct} \\

\midrule

\textbf{\ussimple}~\citep{cen2025large}
& 260 & 106 & 1 & 27,550
& 9.15 & 11,501 \\

\textbf{\jp}~\citep{miyazaki2026ai}
& 25 & 7 & 4 & 650
& 63.1 & 5,750 \\

\bottomrule
\end{tabular}
\end{adjustbox}
\vspace{-4mm}
\end{table}

%
%

\input{fig/channel}

\section{Motivating Insights}
\label{sec:motivating}

Before evaluating rewriting strategies, we first motivate the study of potential risks of \textsc{VoteGEO} by analyzing what types of sources generative engines use and how long the same webpages remain visible across days. We focus on generative engine citations in \jp, with \google results in \ussimple serving as a reference.

\subsection{What Sources Do Generative Engines Use?}
\label{sec:insight-preference}

We group each observed domain into four source channels ordered by how costly it is for a candidate or party to publish or modify content: \textit{owned/self-published}, e.g., candidate or party websites; \textit{open-platform UGC}, e.g., Reddit; \textit{earned/paid press}, e.g., news outlets; and \textit{independent institution}, e.g., government agencies. Specifically, we first use \texttt{claude-opus-5} with web access to verify the organization behind each domain and assign an initial label, followed by manual review of low-confidence and ambiguous cases. The labeling rules and representative examples are provided in~\trref{}.
For each engine on each day, we pool source occurrences across its query cells and compute the \textit{source share} of each channel as its fraction of all source occurrences. For the 106-day U.S. \google reference, we use a 7-day centered moving average to summarize the longer-term pattern. The Japan generative engine series covers only 7 days and is therefore reported as raw daily values.

\stitle{Generative engines often cite sources from lower-cost channels}
Figure~\ref{fig:source-channel-share} shows that \textit{earned/paid press} remains the largest source channel, accounting for 46.4--60.9\% of citations across the four generative engines. At the same time, \textit{owned/self-published} sources account for 24.8--44.3\%, while \textit{open-platform UGC} reaches as high as 22.2\%. For reference, their shares in U.S. \google results are 5.1\% and 4.5\%, respectively, with these cross-setting differences treated descriptively. \textit{Independent institution} sources account for only 1.4--3.0\% of generative-engine citations, compared with 39.3\% in the U.S. \google reference. These source patterns are consistent with prior studies showing that generative engines answering election-related queries cite candidate or party websites, UGC platforms, and news media~\cite{damiao2026evidence,grossman2026generative,schatto2025chatgpt,miyazaki2026ai}.
Source-channel shares also change little across the seven Japan collection days within each generative engine. The longer U.S. \google reference similarly shows a relatively stable source composition over the 106-day window, with \textit{earned/paid press} and \textit{independent institution} remaining dominant. 
These patterns indicate that generative engines frequently use sources that candidates or parties can publish or modify at relatively low cost, raising concerns about the influence of externally controlled content on electoral information.

\subsection{How Long Do the Same Sources Stay Visible?}
\label{sec:insight-stable}

To examine how long the same sources remain visible, we measure their overlap across days. For each query and engine, we define \textit{cross-day source overlap} as the fraction of distinct URLs or domains observed across the sampled days that appear on every sampled day. For \jp, we use all available collection days and group queries by policy issue. For the U.S. \google reference, we use 7 evenly spaced days spanning the 106-day collection window and aggregate results by query subcategory.

\begin{table}[t]
\centering
\small
\caption{Cross-day source overlap at URL and domain levels.}
\label{tab:source_overlap}
\vspace{-2mm}

\setlength{\tabcolsep}{3pt}
\renewcommand{\arraystretch}{1.04}

\resizebox{\columnwidth}{!}{
\begin{tabular}{
@{}
lrr
@{\hspace{6pt}}
lrr
@{}
}
\toprule
\multicolumn{3}{c}{\textbf{U.S.}}
&
\multicolumn{3}{c}{\textbf{Japan}}
\\
\cmidrule(lr){1-3}
\cmidrule(lr){4-6}

\textbf{{Subcategory}}
& \textbf{URL}
& \textbf{Domain}
&
\textbf{Policy issue}
& \textbf{URL}
& \textbf{Domain}
\\
\midrule

Nominees                 & 25.0 & 25.0 & Social insurance          & 8.8 & 21.8 \\
Background               & 12.5 & 14.7 & Nuclear power             & 8.7 & 17.9 \\
Presidential candidates  & 12.1 & 16.4 & Consumption tax           & 7.8 & 19.0 \\
Election outcomes        & 10.9 & 16.7 & Diet seat reduction       & 7.0 & 17.9 \\
Issue importance         & 8.4  & 12.2 & Espionage law             & 6.8 & 17.7 \\
Platform                 & 8.3  & 9.5  & Foreign workers           & 6.6 & 19.4 \\
Evaluation               & 7.1  & 11.1 & Permanent residency       & 6.5 & 21.9 \\
Vice president           & 6.0  & 10.1 & Defense spending          & 6.4 & 17.0 \\
Candidate positions      & 4.0  & 7.5  & Corporate donations       & 6.3 & 15.9 \\
Public opinion           & 3.1  & 7.3  & {Separate surnames}    & 5.8 & 16.0 \\
Comparison               & 0.8  & 2.5  & China relations           & 4.1 & 14.6 \\
Recent stories           & 0.0  & 1.9  & Constitutional amendment  & 2.7 & 14.1 \\

\bottomrule
\end{tabular}
}
\vspace{-4mm}
\end{table}

\stitle{The same sources remain visible longer for more stable topics}
Table~\ref{tab:source_overlap} shows that long-standing policy issues tend to retain more of the same sources across days in generative-engine citations. \textit{Social insurance}, \textit{Nuclear power}, and \textit{Consumption tax} show URL overlap of 7.8--8.8\%, whereas more event-driven issues such as \textit{China relations} and \textit{Constitutional amendment} show lower overlap of 4.1\% and 2.7\%, respectively. Domain overlap is consistently higher than URL overlap, indicating that the same publishers can remain visible even when specific webpages change.
The U.S. \google reference shows a similar within-setting pattern. More stable election-reference queries tend to retain more of the same webpages across days. For example, \textit{Background} queries, such as ``What controversies is Donald Trump the subject of?'', reach 12.5\% URL overlap. In contrast, \textit{Comparison} queries reach 0.8\%, while \textit{Recent stories}, such as ``Are there any recent stories about Kamala Harris that may affect the 2024 US presidential election?'', have no URL shared across all sampled days.
These patterns suggest that rewriting may remain useful longer for stable-topic queries, while current-event queries require fresher content.

\section{GEO Rewriting Strategies}
\label{sec:strategies}

We study 11 rewriting strategies organized into 4 categories. \textit{Structure}, \textit{Evidence}, and \textit{Style} contain 9 strategies adapted from prior GEO research~\citep{geo,c-seo}. We further propose an \textit{Argumentation} category with two strategies for voting scenarios~\citep{lin2025persuading,hackenburg2025levers}.

\subsection{Existing GEO Strategies}
\label{sec:existing-strategies}

\stitle{Structure}
Structure strategies modify page organization and presentation without changing the underlying factual content. Both strategies follow C-SEO~\citep{c-seo}.
\textit{Content Improvement} improves the organization and presentation of the page through clearer sections, more logical ordering, stronger emphasis on key information, and reduced redundancy, while preserving factual information, existing citations, and overall length.
\textit{LLMs.txt} reformats the page into an LLM-oriented layout. It adds an \texttt{LLM Guidance} block with the topic, key facts, and citation priority, reorganizes the body using Markdown headings, key points, summaries, lists, and bold text, and adds a final citation section explaining why the page should be cited and what information it uniquely provides.

\stitle{Evidence}
Evidence strategies strengthen support for existing claims by adding quotations, quantitative evidence, or source attributions. All three follow GEO~\citep{geo}.
\textit{Quotes} inserts direct quotations attributed to credible figures while preserving the remaining content and formatting.
\textit{Statistics} inserts persuasive numerical evidence, such as percentages, growth rates, and counts, at relevant positions while modifying only the sentences receiving the numbers.
\textit{Authoritative} adds five to six relevant attributions to credible sources while preserving the core content, length, and local structure.

\stitle{Style}
Style strategies rewrite the same information at the lexical and sentence levels without adding new substantive content. All four follow GEO~\citep{geo}.
\textit{Fluency} improves sentence flow, clarity, and readability.
\textit{Simple Language} paraphrases sentences using simpler and clearer language while keeping the information and length unchanged.
\textit{Technical Terms} rephrases individual sentences using more technical terminology without adding or removing content.
\textit{Unique Words} introduces more distinctive vocabulary while preserving the original meaning and core information.

\definecolor{nonsig}{gray}{0.68}
\newcommand{\sig}[1]{{#1}}
\newcommand{\ns}[1]{\textcolor{nonsig}{#1}}

\begin{table*}[t]
\centering
\small
\caption{Effect of target specification across strategy categories and engines. Values are mean improvements in percentage points. Black values are statistically significant ($p<0.05$). U = target unspecified, S = target specified.}
\label{tab:rq1-target}
\vspace{-2mm}

\setlength{\tabcolsep}{2.7pt}
\renewcommand{\arraystretch}{1.08}

\begin{tabular}{l*{18}{c}}
\toprule
&
\multicolumn{6}{c}{\textbf{\gptfourmini}} &
\multicolumn{6}{c}{\textbf{{\gptfivemini}}} &
\multicolumn{6}{c}{\textbf{{\gemini}}} \\
\cmidrule(lr){2-7}
\cmidrule(lr){8-13}
\cmidrule(lr){14-19}

\multirow{2}{*}{\textbf{Category}}
&
\multicolumn{2}{c}{\textbf{Mention}}
& \multicolumn{2}{c}{\textbf{SOV}}
& \multicolumn{2}{c}{\textbf{Win Rate}}
&
\multicolumn{2}{c}{\textbf{Mention}}
& \multicolumn{2}{c}{\textbf{SOV}}
& \multicolumn{2}{c}{\textbf{Win Rate}}
&
\multicolumn{2}{c}{\textbf{Mention}}
& \multicolumn{2}{c}{\textbf{SOV}}
& \multicolumn{2}{c}{\textbf{Win Rate}}
\\[-1pt]

&
U & S & U & S & U & S &
U & S & U & S & U & S &
U & S & U & S & U & S \\
\midrule

\textit{Placebo}
& \ns{$-0.18$} & \ns{$-0.09$}
& \ns{$+0.16$} & \ns{$-0.26$}
& \ns{$+0.87$} & \sig{$-1.13$}
& \ns{$-1.40$} & \ns{$-0.44$}
& \ns{$-0.56$} & \ns{$-0.24$}
& \ns{$-1.32$} & \sig{$-0.87$}
& \ns{$+0.50$} & \ns{$+0.29$}
& \ns{$+0.04$} & \ns{$+0.16$}
& \ns{$-0.56$} & \ns{$+0.18$}
\\
\midrule

{Argumentation}
& \ns{$-0.31$} & \sig{$+1.94$}
& \ns{$+0.07$} & \sig{$+1.48$}
& \ns{$+0.86$} & \sig{$+2.57$}
& \ns{$-0.08$} & \ns{$+0.74$}
& \ns{$-0.48$} & \sig{$+0.53$}
& \ns{$-0.93$} & \ns{$+0.52$}
& \ns{$+0.28$} & \sig{$+1.23$}
& \ns{$+0.17$} & \sig{$+1.11$}
& \ns{$+0.84$} & \sig{$+1.58$}
\\

Structure
& \ns{$-0.49$} & \sig{$+2.81$}
& \ns{$+0.19$} & \sig{$+1.30$}
& \ns{$+1.03$} & \sig{$+2.14$}
& \ns{$-1.28$} & \sig{$+1.76$}
& \sig{$-0.87$} & \sig{$+0.89$}
& \ns{$-0.73$} & \sig{$+0.73$}
& \ns{$+0.61$} & \sig{$+1.89$}
& \ns{$+0.25$} & \sig{$+1.47$}
& \ns{$+0.36$} & \sig{$+1.84$}
\\

Evidence
& \ns{$+0.32$} & \sig{$+1.72$}
& \ns{$+0.27$} & \sig{$+0.76$}
& \ns{$+0.94$} & \sig{$+0.91$}
& \ns{$-0.54$} & \ns{$+0.52$}
& \sig{$-0.65$} & \sig{$+0.44$}
& \sig{$-0.92$} & \ns{$+0.14$}
& \ns{$-0.04$} & \ns{$+0.28$}
& \ns{$+0.04$} & \sig{$+0.88$}
& \ns{$+0.31$} & \sig{$+0.90$}
\\

Style
& \ns{$-0.11$} & \ns{$+0.38$}
& \ns{$+0.15$} & \ns{$+0.03$}
& \sig{$+1.46$} & \ns{$-0.21$}
& \ns{$+0.00$} & \ns{$+0.18$}
& \ns{$-0.39$} & \ns{$+0.18$}
& \sig{$-1.08$} & \ns{$-0.17$}
& \ns{$+0.37$} & \ns{$+0.17$}
& \ns{$+0.21$} & \sig{$+0.94$}
& \ns{$+0.59$} & \sig{$+0.52$}
\\

\bottomrule
\end{tabular}

\vspace{1mm}
\end{table*}

\subsection{Argumentation Strategies}
\label{sec:argument-strategies}

Existing GEO strategies primarily modify page structure, supporting evidence, or linguistic style. Voting webpages may additionally contain competing political claims or recommendations. Motivated by political persuasion research on argument framing and counterargument~\citep{lin2025persuading,hackenburg2025levers}, we introduce two argumentation strategies.

\stitle{Comparative Distinction}
This strategy strengthens a comparative case by emphasizing what makes one option distinctively strong relative to plausible alternatives. It foregrounds concrete distinctions in policy positions, commitments, timelines, mechanisms, or other actionable details that can make the option stand out rather than appear merely acceptable. The rewriting prompt instructs the model to preserve the original factual information and maintain a similar overall length. This design is motivated by prior work on contrast advertising in political campaigns, which compares candidates or positions to differentiate the favored option from alternatives~\citep{pfau2002issue}.

\stitle{Refutational}
This strategy strengthens a position through refutational two-sided argumentation~\citep{okeefe1999good,hackenburg2025levers}. It introduces plausible opposing arguments and explicitly rebuts them with reasoning or supporting information. Each opposing argument must be substantive and directly addressed rather than weakened into a straw man. The original core content and overall position are preserved.
\definecolor{nonsig}{gray}{0.68}

\newcommand{\sigres}[1]{{#1}}

\newcommand{\nsres}[1]{%
    \textcolor{nonsig}{#1}%
}

\newcommand{\base}[1]{\textcolor{black!60}{#1}}

\begin{table*}[t]
\centering

\caption{Target-specified rewriting across elections and target tiers, pooled across the three engines. Baseline is mean exposure (\%), and other values are mean improvements ($\Delta$, percentage points). Black values are statistically significant ($p<0.05$).}
\label{tab:rq2-election}

\vspace{-1mm}
\small
\setlength{\tabcolsep}{4.2pt}
\renewcommand{\arraystretch}{1.15}

\begin{tabular}{lcccc cccc cccc}
\toprule
&
\multicolumn{4}{c}{\textbf{Mention Coverage}}
&
\multicolumn{4}{c}{\textbf{Share of Voice}}
&
\multicolumn{4}{c}{\textbf{Win Rate}}
\\
\cmidrule(lr){2-5}
\cmidrule(lr){6-9}
\cmidrule(lr){10-13}

\textbf{Category}
&
\textbf{U.S.} & \textbf{Top} & \textbf{Mid} & \textbf{Low}
&
\textbf{U.S.} & \textbf{Top} & \textbf{Mid} & \textbf{Low}
&
\textbf{U.S.} & \textbf{Top} & \textbf{Mid} & \textbf{Low}
\\

\textit{Baseline}
&
\base{85.86} & \base{72.41} & \base{72.43} & \base{28.55}
&
\base{54.41} & \base{12.03} & \base{10.10} & \base{3.73}
&
\base{58.88} & \base{13.72} & \base{9.25} & \base{2.34}
\\
\midrule

{Argumentation}
&
\sigres{$+0.52$}
& \sigres{$+2.74$}
& \sigres{$+1.75$}
& \sigres{$+1.96$}
&
\sigres{$+1.29$}
& \sigres{$+1.04$}
& \sigres{$+0.69$}
& \sigres{$+0.46$}
&
\sigres{$+1.97$}
& \sigres{$+1.72$}
& \nsres{$+0.74$}
& \nsres{$+0.52$}
\\

Structure
&
\sigres{$+0.78$}
& \sigres{$+4.28$}
& \sigres{$+2.55$}
& \sigres{$+4.12$}
&
\sigres{$+1.54$}
& \sigres{$+1.15$}
& \sigres{$+0.74$}
& \sigres{$+0.83$}
&
\sigres{$+1.94$}
& \sigres{$+1.48$}
& \nsres{$+0.79$}
& \sigres{$+1.20$}
\\

Evidence
&
\nsres{$+0.36$}
& \sigres{$+1.76$}
& \nsres{$+0.74$}
& \sigres{$+1.03$}
&
\sigres{$+1.09$}
& \sigres{$+0.35$}
& \sigres{$+0.28$}
& \sigres{$+0.20$}
&
\sigres{$+1.21$}
& \nsres{$+0.20$}
& \nsres{$-0.25$}
& \nsres{$+0.15$}
\\

Style
&
\nsres{$+0.08$}
& \sigres{$+1.27$}
& \nsres{$-0.26$}
& \nsres{$+0.21$}
&
\sigres{$+0.83$}
& \nsres{$+0.09$}
& \nsres{$+0.01$}
& \nsres{$-0.00$}
&
\sigres{$+0.45$}
& \nsres{$-0.46$}
& \nsres{$-0.48$}
& \nsres{$+0.09$}
\\

\bottomrule
\end{tabular}

\end{table*}

\section{\ourbench}
\label{sec:exp}

This section evaluates \textsc{VoteGEO} through three research questions:

\begin{itemize}[topsep=2pt,itemsep=1pt,parsep=0pt,partopsep=0pt,leftmargin=11pt]
    \item \textbf{RQ1: Does Target Specification Matter?}
    We compare rewriting with and without an explicitly specified target party.

    \item \textbf{RQ2: Which Rewriting Strategies Work?}
    We compare the 11 rewriting strategies across engines and elections.

    \item \textbf{RQ3: How Do Voter Personas Affect GEO?}
    We examine how voter personas affect baseline exposure and rewriting gains.
\end{itemize}
We report the main results below and provide full results across the evaluated settings in~\trref{}.

\subsection{Benchmark Setup}
\label{sec:benchmark-setup}

\stitle{Datasets}
We retain 109 U.S. queries from 5 subcategories that directly evaluate or compare candidates: \textit{Candidate positions} (51), \textit{Comparison} (29), \textit{Public opinion} (17), \textit{Evaluation} (9), and \textit{Vice president} (3). For each query, we independently sample 7 collection days uniformly at random from its available days using a fixed seed. Together, the selected query cells cover the full 106-day collection window. For Japan, we use all 25 queries and all available collection days.
We use \gptfourmini, \gptfivemini, and \gemini as generative engines from OpenAI and Google. Following the fixed-pool GEO setting in Section~\ref{sec:formulation}, web search is disabled and responses are generated from the provided webpage pools. For the U.S., all three engines use the same pools constructed from historical \google results. For Japan, \gptfourmini and \gptfivemini use their corresponding GPT pools. Because the original Grok models are no longer available, \gemini is evaluated separately on both Grok webpage pools, and the two sets of results are combined for reporting. All engines use temperature 0. This yields 2,289 U.S. and 650 Japan query cells, for 2,939 in total.

\stitle{Benchmark Procedure}
We define targets at the party level to keep the U.S. and Japan settings comparable, with candidate mentions attributed to their corresponding parties. For the U.S., we use the \textit{Democratic} Party as the primary target. Results using the \textit{Republican} Party as the target show the same overall pattern. For Japan, we divide the 11 parties into \textit{Top}, \textit{Mid}, and \textit{Low} tiers of sizes 4, 4, and 3 based on final proportional-representation vote share. For each webpage pool, we randomly sample one mentioned party from each tier as the target and then randomly sample one webpage mentioning that target as $d_i$. The target and $d_i$ remain fixed across replications.
For each target and rewriting strategy, we use \gptfourmini to rewrite $d_i$ and generate a \textit{treatment} response by replacing only $d_i$ with its rewritten version in the corresponding webpage pool. The same baseline and placebo responses are shared across targets within each query cell.
We repeat the procedure three times, keeping the target and $d_i$ fixed while independently regenerating rewrites and responses and shuffling the input order.

\stitle{Exposure Metrics}
We instantiate the exposure function $\phi(t,y)$ in Equation~\ref{eq:obj} using three metrics. For each sentence, its word count is divided equally among all parties mentioned in that sentence, while sentences mentioning no party contribute to none. \textit{Mention Coverage} is the fraction of responses that mention the target party. \textit{Share of Voice} (SOV) is the average fraction of party-attributed words assigned to the target. \textit{Win Rate} is the fraction of responses in which the target receives the largest attributed word count, with ties receiving equal fractional credit.
All three metrics are set to zero when no party is mentioned. For each benchmark instance, we compute treatment-minus-baseline within each replication and then average over the three replications. We report the absolute improvement in percentage points. Placebo-minus-baseline is computed in the same way to quantify run-to-run variation. Statistical significance is assessed using two-sided paired $t$-tests over benchmark instances.

\subsection{RQ1: Does Target Specification Matter?}

We first examine whether explicitly specifying the target affects rewriting effectiveness. Compared with the target-unspecified setting, the target-specified setting explicitly names the target party in the rewriting prompt, while the remaining setup is unchanged. Table~\ref{tab:rq1-target} reports the mean exposure improvement across strategy categories and generative engines, together with the placebo as the no-rewriting reference.

\stitle{Target specification is important for effective rewriting}
Target-unspecified rewriting rarely increases target exposure, whereas target-specified rewriting does so in most evaluated settings. Among the 36 category, engine, and metric settings, target-unspecified rewriting produces only one significant positive gain, while four show significant negative changes and the remaining 31 are not significant. In contrast, target-specified rewriting produces significant positive gains in 24 of 36 settings, including 10 of 12 for SOV. Positive gains appear across all three engines and the Structure, Evidence, and Argumentation categories, while Style is less consistent.
The difference arises because election webpages often discuss multiple competing parties, whereas existing GEO settings typically involve webpages centered on a single entity. In the U.S., 92\% of webpages in the pools mention both major parties, making the intended target ambiguous without explicit specification. A failure analysis of target-unspecified Comparative Distinction rewrites in Japan shows the same issue. About one quarter strengthen an opposing party, while 25\% remove the target name. We therefore use target-specified rewriting in all subsequent experiments.

\input{fig/strategy-rank}

\begin{table*}[t]
\centering

\caption{Effect of voter personas on baseline SOV and target-specified rewriting in the U.S. Base is baseline SOV, $\Delta$ is the rewriting gain, and Mod is its difference from the persona-free gain. Black values are significant ($p<0.05$). CI = Content Improvement; CD = Comparative Distinction.}
\label{tab:rq3-persona}

\vspace{-2mm}
\small
\setlength{\tabcolsep}{2.2pt}
\renewcommand{\arraystretch}{1.08}

\begin{tabular}{ll rrrrr rrrrr rrrrr}
\toprule
&
&
\multicolumn{5}{c}{\textbf{\gptfourmini}}
&
\multicolumn{5}{c}{\textbf{{\gptfivemini}}}
&
\multicolumn{5}{c}{\textbf{{\gemini}}}
\\
\cmidrule(lr){3-7}
\cmidrule(lr){8-12}
\cmidrule(lr){13-17}

&
&
&
\multicolumn{2}{c}{\textbf{$\Delta$}}
&
\multicolumn{2}{c}{\textbf{Mod}}
&
&
\multicolumn{2}{c}{\textbf{$\Delta$}}
&
\multicolumn{2}{c}{\textbf{Mod}}
&
&
\multicolumn{2}{c}{\textbf{$\Delta$}}
&
\multicolumn{2}{c}{\textbf{Mod}}
\\[-1pt]

\textbf{{Category}}
&
\textbf{Persona}
&
\textbf{Base} & \textbf{CI} & \textbf{CD} & \textbf{CI} & \textbf{CD}
&
\textbf{Base} & \textbf{CI} & \textbf{CD} & \textbf{CI} & \textbf{CD}
&
\textbf{Base} & \textbf{CI} & \textbf{CD} & \textbf{CI} & \textbf{CD}
\\
\midrule

--
&
\textit{{Persona-free}}
&
56.5 & \sig{+0.70} & \sig{+1.42} & -- & --
&
50.1 & \sig{+1.14} & \sig{+1.81} & -- & --
&
56.0 & \ns{+0.30} & \sig{+1.53} & -- & --
\\

\midrule

\multirow{9}{*}{
    \shortstack[l]{
        \textit{Demographic}\\
        \textit{identity}~\cite{cen2025large}
    }
}
&
White
&
55.7 & \sig{+1.54} & \sig{+2.45} & \sig{+0.84} & \sig{+1.03}
&
51.5 & \ns{+0.89} & \ns{+0.76} & \ns{$-$0.25} & \ns{$-$1.06}
&
56.3 & \ns{+0.57} & \sig{+1.41} & \ns{+0.27} & \ns{$-$0.13}
\\

&
Black
&
58.0 & \ns{+0.56} & \sig{+2.04} & \ns{$-$0.14} & \ns{+0.63}
&
52.2 & \sig{+1.38} & \ns{+0.59} & \ns{+0.24} & \ns{$-$1.23}
&
57.2 & \ns{+0.59} & \sig{+1.78} & \ns{+0.29} & \ns{+0.25}
\\

&
Hispanic
&
57.6 & \sig{+0.65} & \sig{+1.77} & \ns{$-$0.05} & \ns{+0.36}
&
51.8 & \sig{+1.41} & \sig{+1.50} & \ns{+0.27} & \ns{$-$0.32}
&
56.3 & \sig{+1.02} & \sig{+2.25} & \ns{+0.72} & \ns{+0.72}
\\

&
Asian
&
57.2 & \sig{+0.91} & \sig{+2.09} & \ns{+0.20} & \ns{+0.67}
&
51.9 & \sig{+1.19} & \ns{+0.53} & \ns{+0.05} & \sig{$-$1.29}
&
56.5 & \sig{+1.30} & \sig{+1.89} & \sig{+1.00} & \ns{+0.36}
\\

&
American Indian
&
56.9 & \sig{+1.09} & \sig{+1.90} & \ns{+0.39} & \ns{+0.49}
&
52.6 & \ns{+0.46} & \ns{+0.16} & \ns{$-$0.68} & \sig{$-$1.65}
&
56.5 & \ns{+0.64} & \sig{+1.58} & \ns{+0.34} & \ns{+0.04}
\\

&
Pacific Islander
&
57.1 & \sig{+0.71} & \sig{+1.74} & \ns{+0.01} & \ns{+0.33}
&
52.3 & \sig{+1.30} & \sig{+1.14} & \ns{+0.16} & \ns{$-$0.68}
&
56.4 & \sig{+1.26} & \sig{+2.10} & \sig{+0.96} & \ns{+0.57}
\\

&
Man
&
56.1 & \sig{+0.95} & \sig{+1.44} & \ns{+0.24} & \ns{+0.02}
&
51.4 & \sig{+1.01} & \sig{+1.64} & \ns{$-$0.13} & \ns{$-$0.18}
&
56.2 & \sig{+0.79} & \sig{+1.61} & \ns{+0.49} & \ns{+0.08}
\\

&
Woman
&
57.4 & \sig{+1.02} & \sig{+2.10} & \ns{+0.32} & \ns{+0.68}
&
52.0 & \sig{+1.15} & \sig{+1.14} & \ns{+0.01} & \ns{$-$0.67}
&
56.5 & \ns{+0.68} & \sig{+1.72} & \ns{+0.38} & \ns{+0.19}
\\

&
Non-binary
&
56.8 & \sig{+0.89} & \sig{+1.74} & \ns{+0.18} & \ns{+0.32}
&
51.7 & \ns{$-$0.22} & \ns{+0.58} & \ns{$-$1.36} & \sig{$-$1.23}
&
56.5 & \sig{+0.90} & \sig{+1.33} & \ns{+0.60} & \ns{$-$0.20}
\\

\midrule

\multirow{3}{*}{
    \shortstack[l]{
        \textit{Partisan}\\
        \textit{identity}~\cite{cen2025large}
    }
}
&
Democrat
&
61.1 & \ns{+0.41} & \sig{+2.20} & \ns{$-$0.30} & \ns{+0.78}
&
54.3 & \sig{+1.08} & \sig{+1.76} & \ns{$-$0.06} & \ns{$-$0.06}
&
57.8 & \sig{+0.76} & \sig{+2.16} & \ns{+0.46} & \ns{+0.63}
\\

&
Independent
&
56.8 & \ns{+0.49} & \sig{+1.57} & \ns{$-$0.22} & \ns{+0.15}
&
53.3 & \ns{+0.80} & \ns{+0.69} & \ns{$-$0.34} & \ns{$-$1.12}
&
56.9 & \ns{+0.30} & \sig{+1.64} & \ns{+0.00} & \ns{+0.11}
\\

&
Republican
&
49.5 & \ns{+0.27} & \sig{+1.36} & \ns{$-$0.43} & \ns{$-$0.05}
&
50.9 & \sig{+0.95} & \ns{+0.78} & \ns{$-$0.19} & \ns{$-$1.04}
&
54.8 & \sig{+0.84} & \sig{+1.21} & \ns{+0.54} & \ns{$-$0.32}
\\

\midrule

\multirow{3}{*}{
    \shortstack[l]{
        \textit{{Questionnaire-derived}}\\
        \textit{{profile}}~\cite{lin2025persuading}
    }
}
&
Lean left
&
56.9 & \ns{+0.45} & \sig{+1.45} & \ns{$-$0.25} & \ns{+0.04}
&
54.2 & \sig{+0.95} & \sig{+1.11} & \ns{$-$0.19} & \ns{$-$0.71}
&
57.1 & \ns{+0.22} & \sig{+1.15} & \ns{$-$0.08} & \ns{$-$0.38}
\\

&
Swing
&
57.1 & \ns{$-$0.05} & \sig{+0.86} & \ns{$-$0.75} & \ns{$-$0.56}
&
54.1 & \ns{+0.91} & \ns{+0.72} & \ns{$-$0.23} & \ns{$-$1.10}
&
56.8 & \sig{+0.77} & \sig{+1.25} & \ns{+0.47} & \ns{$-$0.28}
\\

&
Lean right
&
56.9 & \sig{+0.80} & \sig{+1.50} & \ns{+0.09} & \ns{+0.09}
&
53.4 & \sig{+0.97} & \sig{+1.36} & \ns{$-$0.17} & \ns{$-$0.46}
&
56.1 & \sig{+0.92} & \sig{+1.82} & \ns{+0.62} & \ns{+0.29}
\\

\bottomrule
\end{tabular}

\vspace{1mm}
\end{table*}

\subsection{RQ2: Which Rewriting Strategies Work?}

We next compare rewriting effectiveness across generative engines and elections, and then examine individual strategies.

\stitle{Structure and Argumentation show stronger gains than Evidence and Style across engines}
Table~\ref{tab:rq1-target} shows that Structure produces significant gains in all 9 engine and metric settings and Argumentation in 7, compared with 6 for Evidence and only 2 for Style. Notably, Structure consistently improves all three exposure metrics across engines, while Argumentation also shows positive gains in most settings. The magnitude of improvement varies across engines, with smaller gains on \gptfivemini. This coincides with its higher refusal rate on evaluative queries such as ``which candidate is most X'', leaving fewer responses for rewriting to affect.

\stitle{Structure and Argumentation remain strongest across elections, with larger relative gains for lower-exposure parties}
Table~\ref{tab:rq2-election} shows that Structure and Argumentation show stronger gains than Evidence and Style in both the U.S. and Japan. Across elections and target tiers, rewriting produces similar absolute gains, but these gains correspond to larger relative improvements for parties with lower baseline exposure. For example, under Structure in Japan, the Top and Low tiers gain 4.28 and 4.12 percentage points in Mention Coverage, corresponding to relative improvements of 5.9\% and 14.4\%. For Win Rate, gains of 1.48 and 1.20 points correspond to relative improvements of 10.8\% and 51.1\%. Thus, although the absolute exposure gains are modest, comparable gains can represent much larger relative improvements for parties with low baseline.

\stitle{Content Improvement and Comparative Distinction perform best overall across settings}
Figure~\ref{fig:rq2-strategy-rank} compares the 11 strategies across 18 combinations of engine, election, and exposure metric. Within each setting, strategies are ranked by improvement over the baseline. The bars show how often each strategy receives each rank, while Mean and Top-3 report its average rank and number of Top-3 finishes. Content Improvement and Comparative Distinction perform best overall, with mean ranks of 2.61 and 2.67 and Top-3 finishes in 13 and 14 of 18 settings, respectively. Their strong performance aligns with the category-level finding that Structure and Argumentation show the strongest gains.

\subsection{RQ3: How Do Voter Personas Affect GEO?}
\label{sec:rq3}

We finally examine how voter personas affect baseline exposure and rewriting gains. Within each query cell, we keep the target and rewritten webpage fixed and vary only the persona description prepended to the generation prompt. We evaluate Content Improvement and Comparative Distinction under 15 U.S. personas: 9 demographic identities, 3 partisan identities, and 3 questionnaire-derived profiles. We focus on Share of Voice (SOV). For persona $p$, $\Delta_p$ denotes the rewriting gain and $\mathrm{Mod}_p=\Delta_p-\Delta_{\emptyset}$ its difference from the persona-free gain. Mention Coverage and Win Rate show the same overall patterns.

\stitle{Rewriting remains effective across personas, with no consistent increase or decrease in gains}
Table~\ref{tab:rq3-persona} shows that demographic identities shift baseline SOV by $-0.8$ to $+2.5$ percentage points, whereas partisan identities shift it by $-7.0$ to $+4.6$ points. This is consistent with prior findings that explicit political information has a larger effect on election-related responses than demographic information~\cite{cen2025large,miyazaki2026ai}.  Rewriting nevertheless remains effective across personas. Comparative Distinction yields significant positive SOV gains under all 15 personas on both \gptfourmini and \gemini, while 29 of the 30 persona--strategy gains are positive on \gptfivemini. We also evaluate gender personas in Japan, the only demographic dimension shared with the U.S. dataset, and observe consistent results. In addition, 7 of the 90 Mod estimates are nominally significant, but none remain significant after clustering by base query and Benjamini–Hochberg correction (See~\trref{}). These effects have mixed directions across engines. Thus, rewriting gains remain broadly positive across personas, while persona conditioning shows no consistent tendency to increase or decrease these gains.

\stitle{Partisan identity shifts baseline exposure more than rewriting a single webpage}
Across all engines, baseline Democratic SOV follows the same \textit{Democrat $>$ Independent $>$ Republican} ordering, showing that explicit voter partisanship systematically shifts Democratic exposure. The Democrat--Republican baseline gap reaches $11.6$, $3.4$, and $3.0$ percentage points on \gptfourmini, \gptfivemini, and \gemini, respectively. For comparison, Comparative Distinction increases Democratic SOV by only $2.20$, $1.76$, and $2.16$ points under the Democrat persona. Thus, explicit partisan identity can shift baseline exposure more than rewriting a single webpage. The Independent persona produces higher Democratic SOV than persona-free generation across all three engines, consistent with prior evidence of left-leaning tendencies in default LLM outputs in U.S. political contexts~\cite{kim2026polar}. Despite these baseline shifts, rewriting remains effective across all three partisan identities.

\stitle{Baseline shifts from richer voter profiles vary across engines}
We use a shallow decision tree to identify three interpretable, preference-enriched subgroups from \usrich and construct questionnaire-derived profiles from their defining attributes. These selected profiles are not intended to represent all voters in their respective preference groups. We exclude explicit ideology and candidate-evaluation variables and use voting-lean labels only for profile construction, not in the generation prompts. For reporting, we refer to the profiles as \textit{Lean left}, \textit{Swing}, and \textit{Lean right}. Their prompts describe, respectively, a voter with high political interest, low religiosity, and strong emphasis on candidate personality, a voter with low political interest, lower emphasis on policy positions, and moderate emphasis on candidate personality, and a voter with high political interest, high religiosity, and low emphasis on candidate personality. Further construction details are provided in~\trref{}.
Table~\ref{tab:rq3-persona} shows clear differences across engines. On \gptfourmini, profile-conditioned baselines remain close to persona-free generation and differ from their corresponding partisan baselines by up to $7.4$ percentage points. The maximum gaps shrink to $2.5$ points on \gptfivemini and $1.3$ points on \gemini. Thus, baseline exposure shifts vary with how strongly each engine responds to the partisan information implicit in these profiles. \gptfivemini and \gemini show stronger shifts in this direction than \gptfourmini. Rewriting remains effective across these profiles, with positive gains in 17 of the 18 engine--profile--strategy combinations and significant positive gains from Comparative Distinction in 8 of 9 combinations.
\section{Conclusion}
In this work, we introduce \ourbench to investigate the potential risks of GEO in voting scenarios. We find that generative engines frequently cite lower-cost source channels and that some webpages remain visible across days, motivating our subsequent evaluation of how webpage rewriting affects electoral exposure.
Our experiments show that explicitly specifying the target is important for effective rewriting, while Structure and Argumentation strategies show stronger gains across engines and elections. We further find that rewriting gains remain broadly positive across voter personas, while partisan information can strongly shift baseline exposure.
Future work may explore ethically controlled end-to-end evaluation and further investigate contextual sentiment, factual quality, semantic fidelity, and human persuasion.

\section{Limitations and Ethics Statement}

\stitle{Offline evaluation}
Our rewriting experiments use fixed historical webpage pools with web retrieval disabled, leaving the effects of rewriting on live retrieval and citation selection unexamined. A live end-to-end evaluation would typically require publishing rewritten political content for crawling and indexing by search and generative engines. Even after removal, such content could persist in caches or archived snapshots and potentially reach real voters. To avoid these risks, we conduct all rewriting experiments offline without publishing rewritten political content to live websites.

\stitle{Exposure and factuality}
Our evaluation measures changes in candidate or party exposure without distinguishing favorable from unfavorable mentions or assessing factual accuracy and voter persuasion. Moreover, we do not systematically verify the factual correctness or semantic fidelity of rewritten webpages. Some established GEO Evidence strategies may introduce supporting material not present in the original webpage, while our Structure and Argumentation prompts instruct the rewriter to preserve the original factual information or core position. We therefore interpret observed gains only as changes in exposure, rather than improvements in information quality or evidence of voter persuasion.

\stitle{Privacy and dual-use risks}
The persona analysis uses previously collected datasets and reports aggregate results without targeting individual voters. Nevertheless, releasing rewriting strategies and their effectiveness comparisons carries dual-use risks, as these findings could be used to influence political exposure in deployed systems. We hope these findings help researchers understand the effectiveness and risks of GEO and inform safeguards against manipulation through retrieved web content.


\begin{acks}
We thank Ying Kiat Tan and Tsuhan Chen for their valuable discussions and constructive feedback on this work.
\end{acks}

\bibliographystyle{ACM-Reference-Format}
\bibliography{refs}

\appendix
\section{Appendix}
\label{appendix}


\begin{table*}[!t]
\centering
\small
\caption{Composition of the 260 U.S. base queries retained from \citet{cen2025large}. $\ast$ marks the five subcategories used in Section~\ref{sec:benchmark-setup}.}
\label{tab:us_query_composition}
\vspace{-2mm}
\setlength{\tabcolsep}{3pt}
\renewcommand{\arraystretch}{1.02}

\begin{tabular}{lllr}
\toprule
\textbf{Category}
& \textbf{Subcategory}
& \textbf{Example query}
& \textbf{\#} \\
\midrule

\textit{Predictions}
& Election outcomes
& How will South Carolina vote in the 2024 US presidential election?
& 105 \\

\textit{Election issues}
& Candidate positions$^\ast$
& What is Kamala Harris's position on immigration?
& 51 \\

\textit{Candidates}
& Comparison$^\ast$
& Which 2024 US presidential candidate has the best chance of winning?
& 29 \\

\textit{Election issues}
& Public opinion$^\ast$
& Which candidate do voters believe will handle immigration best?
& 17 \\

\textit{Issues}
& Issue importance
& What is the importance of and debate around the economy?
& 17 \\

\textit{Candidates}
& Background
& What controversies is Donald Trump the subject of?
& 12 \\

\textit{Candidates}
& Evaluation$^\ast$
& What are Donald Trump's greatest strengths and weaknesses?
& 9 \\

\textit{Candidates}
& Presidential candidates
& Can you describe Joe Biden's voting record before 2024?
& 9 \\

\textit{Candidates}
& Platform
& What is Kamala Harris's 2024 political platform?
& 3 \\

\textit{Candidates}
& Vice president$^\ast$
& How will the 2024 VP picks affect the election?
& 3 \\

\textit{Events}
& Recent stories
& Are there recent stories about Donald Trump that may affect the election?
& 3 \\

\textit{Candidates}
& Nominees
& Who are the 2024 US presidential nominees?
& 2 \\

\midrule
\textbf{Total} & & & \textbf{260} \\
\bottomrule
\end{tabular}
\end{table*}

\subsection{Data Preprocessing}
\label{appendix-data}

\stitle{Query cell sampling}
We retain 260 U.S. base queries from \citet{cen2025large} that are relevant to voter decision making and the presidential election. The rewriting benchmark uses 109 queries from five subcategories: \textit{Candidate positions} (51), \textit{Comparison} (29), \textit{Public opinion} (17), \textit{Evaluation} (9), and \textit{Vice president} (3). To sample the election period without tying all queries to the same calendar dates, we independently select seven available collection days for each query using a fixed seed, jointly covering the full 106-day window. The Japanese set contains 12 policy issues with two opposing stance conditions plus one query without a policy stance, and we use all available collection days.
The cross-day source-overlap analysis instead requires a common time span across queries, so all U.S. queries use the same seven evenly spaced dates: July 23, August 10, August 27, September 13, October 1, October 19, and November 5, 2024. The Japan overlap analysis uses all available days for each source engine.

\stitle{Content retrieval details}
We attempt retrieval for 11,895 U.S. and 6,202 Japanese URLs. When a URL appears on multiple dates, we use its earliest observed date, and URLs are normalized by resolving archive wrappers and removing tracking parameters and fragments before deduplication. Main text is extracted using \texttt{trafilatura}, with \texttt{readability-lxml} as a fallback. The resulting success rates are 96.7\% for the U.S. and 92.7\% for Japan.

\stitle{Source-Channel Labeling}
We label sources at the domain level using the four channels defined in Section~\ref{sec:insight-preference}. Representative examples are \texttt{kamalaharris.com} and \texttt{jimin.jp} for \textit{owned/self-published}, \texttt{reddit.com} and \texttt{note.com} for \textit{open-platform UGC}, \texttt{nytimes.com} and \texttt{asahi.com} for \textit{earned/paid press}, and \texttt{fec.gov} and \texttt{soumu.go.jp} for \textit{independent institution}. Labels follow the organization controlling the domain rather than the type of an individual page.
We use \texttt{claude-opus-5} with web access for initial labeling and manually review low-confidence and ambiguous cases. Source shares count repeated occurrences separately. We report daily shares for \jp and a centered seven-day moving average for the longer \ussimple series, with the main-text percentages averaged over each observation window. The complete domain labels are released with the artifact.

\subsection{Party attribution}
\label{sec:app-benchmark-details}

The response prompt asks engines to wrap candidate and party mentions in \texttt{<ENT>} tags without revealing the closed entity set or target. Tagged spans are normalized and mapped to party entities using a curated longest-match alias dictionary. U.S. aliases for Kamala Harris, Joe Biden, and the Democratic Party map to \textit{Democratic}, while aliases for Donald Trump and the Republican Party map to \textit{Republican}. Running mates and third-party candidates are not mapped. In Japan, the dictionary contains the 11 party names and unambiguous abbreviations, with ambiguous short forms excluded and predecessor aliases mapped to their corresponding successor party. Japanese candidate names are not mapped.
This leaves 0.8\% of tagged spans in baseline Japanese responses and 7.5\% in U.S. responses unmapped, with the latter mostly consisting of third-party candidates. A sentence-level alias scan additionally recovers untagged party mentions using word-boundary matching for ASCII aliases and substring matching for CJK aliases. The same mapping procedure is used for target selection and exposure measurement. Complete alias dictionaries are released with the artifact.

\begin{table}[!t]
\centering
\caption{Effect of gender personas on SOV and rewriting gains across three party tiers in Japan (\texttt{gpt-4o-mini}).}
\label{tab:rq3-jp-gender}

\vspace{-2mm}
\small
\setlength{\tabcolsep}{3pt}
\renewcommand{\arraystretch}{1.08}

\begin{tabular}{ll rrrrr}
\toprule
&
&
&
\multicolumn{2}{c}{\textbf{$\Delta$}}
&
\multicolumn{2}{c}{\textbf{Mod}}
\\[-1pt]

\textbf{Tier}
&
\textbf{Persona}
&
\textbf{Base}
&
\textbf{CI} & \textbf{CD}
&
\textbf{CI} & \textbf{CD}
\\
\midrule

\multirow{3}{*}{Top}
&
\textit{Persona-free}
&
16.7 & \sig{+2.64} & \sig{+3.67} & -- & --
\\

&
Man
&
15.7 & \sig{+4.88} & \sig{+5.76}
& \ns{+2.25} & \ns{+2.10}
\\

&
Woman
&
15.9 & \sig{+4.58} & \sig{+5.23}
& \ns{+1.95} & \ns{+1.57}
\\

\midrule

\multirow{3}{*}{Mid}
&
\textit{Persona-free}
&
7.5 & \ns{+1.17} & \sig{+3.26} & -- & --
\\

&
Man
&
7.4 & \sig{+1.90} & \sig{+2.76}
& \ns{+0.72} & \ns{$-$0.50}
\\

&
Woman
&
6.7 & \ns{+1.77} & \sig{+3.02}
& \ns{+0.60} & \ns{$-$0.24}
\\

\midrule

\multirow{3}{*}{Low}
&
\textit{Persona-free}
&
1.9 & \sig{+1.13} & \ns{+0.99} & -- & --
\\

&
Man
&
1.3 & \ns{+1.25} & \sig{+2.32}
& \ns{+0.12} & \ns{+1.32}
\\

&
Woman
&
1.6 & \ns{+1.28} & \sig{+1.97}
& \ns{+0.15} & \ns{+0.98}
\\

\bottomrule
\end{tabular}
\end{table}

\subsection{Questionnaire-Derived Profile Construction}
\label{sec:rich-persona-construction}

\usrich contains pre-treatment questionnaire responses from 2,306 participants~\citep{lin2025persuading}. We use demographic, political-engagement, and voting-criteria attributes for subgroup discovery, while excluding ideology and candidate-evaluation variables because they directly encode or are strongly associated with political preference.
We use \texttt{lean\_bidentrump\_1} only as a subgroup-discovery label. Scores below 40, from 40 to 60, and above 60 define lean-left, swing, and lean-right groups containing 1,172, 203, and 931 respondents, respectively.
We use decision trees for interpretable subgroup discovery rather than voter classification. Our goal is to identify attribute-defined subgroups enriched for each preference group, rather than to represent all voters in that group. We train balanced multiclass decision trees over Gini and entropy criteria, maximum depths 2--6, and minimum leaf sizes 10--100. For each setting, we select the highest-purity leaf for each preference group subject to at least 20 respondents, then choose the setting maximizing the minimum purity across the three leaves, preferring shallower trees when tied. The selected tree uses entropy, depth 4, and minimum leaf size 20, with five-fold balanced accuracy $0.529 \pm 0.023$ and macro-F1 $0.442 \pm 0.042$. These classification scores are not the primary selection objectives.
For each preference group, GPT-5.5 converts the selected decision rule into a concise voter profile using only the selected attributes and conditions, without the group label, candidate-preference score, or subgroup statistics.
The lean-left subgroup contains 113 respondents with composition 102 / 0 / 11 and follows political interest $>66.5$, religiosity $\leq1.5$, and candidate personality importance $>77.5$. The swing subgroup contains 31 respondents with composition 4 / 21 / 6 and follows political interest $\leq66.5$, policy importance $\leq59.5$, and candidate personality importance between 47.5 and 62.5. The lean-right subgroup contains 66 respondents with composition 5 / 0 / 61 and follows political interest $>66.5$, religiosity $>6.5$, and candidate personality importance $\leq36.5$.
These rules are represented in the generation prompts as a voter with high political interest, low religiosity, and strong emphasis on candidate personality, a voter with lower political interest, lower emphasis on policy positions, and moderate emphasis on candidate personality, and a voter with high political interest, high religiosity, and low emphasis on candidate personality, respectively. The labels \textit{Lean left}, \textit{Swing}, and \textit{Lean right} are used only for reporting. These selected profiles describe high-purity subgroups within the sample and are not intended to represent their entire preference groups.

\begin{table*}[!t]
\centering
\caption{Full target-specified rewriting results across strategies, elections, and response engines. Baseline reports mean exposure (\%), and all other entries report mean paired changes from baseline in percentage points. Black values are statistically significant ($p<0.05$), while gray values are not. Placebo compares two generations without rewriting. For Japan, Top, Mid, and Low denote target-party tiers, and Gemini pools results over the \texttt{grok-4-1-fast} and \texttt{grok-with-x} webpage pools. 4o-mini, 5-mini, and Gemini denote \texttt{gpt-4o-mini}, \texttt{gpt-5-mini}, and \texttt{gemini-2.5-flash}, respectively.}
\label{tab:full-strategy-election-engine}
\vspace{-2mm}

\footnotesize
\setlength{\tabcolsep}{2.6pt}
\renewcommand{\arraystretch}{1.08}

\begin{tabular}{llrrrrrrrrrrrr}
\toprule
& &
\multicolumn{3}{c}{\textbf{U.S.}} &
\multicolumn{3}{c}{\textbf{JP -- Top}} &
\multicolumn{3}{c}{\textbf{JP -- Mid}} &
\multicolumn{3}{c}{\textbf{JP -- Low}} \\
\cmidrule(lr){3-5}
\cmidrule(lr){6-8}
\cmidrule(lr){9-11}
\cmidrule(lr){12-14}

\textbf{Category}
& \textbf{Strategy}
& \textbf{4o-mini} & \textbf{5-mini} & \textbf{Gemini}
& \textbf{4o-mini} & \textbf{5-mini} & \textbf{Gemini}
& \textbf{4o-mini} & \textbf{5-mini} & \textbf{Gemini}
& \textbf{4o-mini} & \textbf{5-mini} & \textbf{Gemini} \\
\midrule

\multicolumn{14}{l}{\textit{Mention Coverage} ($\Delta$, pp)} \\
\midrule

& \textit{Baseline (\%)}
& 90.3 & 74.4 & 92.8
& 51.5 & 65.7 & 83.2
& 25.5 & 75.0 & 87.6
& 4.3 & 24.6 & 39.0 \\

& \textit{Placebo}
& \ns{+0.00} & \ns{$-$0.39} & \ns{$-$0.13}
& \ns{$-$0.27} & \ns{$-$0.19} & \ns{+0.76}
& \ns{$-$1.35} & \ns{$-$3.24} & \ns{+0.86}
& \ns{+0.81} & \ns{+1.91} & \ns{+0.19} \\

\addlinespace[2pt]

\multirow{2}{*}{Argumentation}
& Comparative Distinction
& \ns{+0.83} & \ns{+1.00} & \ns{+0.74}
& \sig{+9.60} & \ns{+1.52} & \ns{+1.81}
& \sig{+7.86} & \ns{+1.33} & \ns{+0.57}
& \sig{+6.78} & \ns{+0.38} & \ns{+1.14} \\

& Refutational
& \ns{$-$0.04} & \ns{$-$0.09} & \ns{+0.70}
& \ns{+4.00} & \ns{+1.33} & \ns{+2.10}
& \ns{+0.27} & \ns{+0.57} & \sig{+2.10}
& \ns{+2.17} & \ns{+1.71} & \sig{+1.91} \\

\addlinespace[2pt]

\multirow{2}{*}{Structure}
& Content Improvement
& \sig{+1.09} & \sig{+1.92} & \ns{$-$0.04}
& \sig{+6.93} & \ns{+2.10} & \sig{+2.57}
& \sig{+5.42} & \ns{+0.76} & \ns{+0.67}
& \sig{+3.79} & \ns{+2.86} & \ns{+1.91} \\

& LLMs.txt
& \ns{+0.39} & \ns{+0.04} & \sig{+1.27}
& \sig{+9.87} & \ns{+3.81} & \sig{+4.38}
& \sig{+9.21} & \ns{+3.62} & \ns{+1.43}
& \sig{+7.05} & \sig{+4.19} & \sig{+6.00} \\

\addlinespace[2pt]

\multirow{3}{*}{Evidence}
& Quotes
& \ns{+0.79} & \ns{+0.17} & \ns{+0.35}
& \sig{+7.47} & \ns{+0.95} & \ns{+1.43}
& \sig{+7.05} & \ns{$-$0.57} & \ns{$-$0.19}
& \sig{+3.25} & \ns{+2.67} & \ns{+0.67} \\

& Statistics
& \ns{+0.79} & \ns{+1.22} & \ns{$-$0.26}
& \ns{+2.93} & \ns{$-$1.14} & \ns{+1.24}
& \sig{+3.79} & \ns{$-$0.95} & \ns{$-$0.86}
& \ns{+1.63} & \ns{+1.33} & \ns{+0.29} \\

& Authoritative
& \ns{+0.35} & \ns{+0.39} & \ns{$-$0.52}
& \sig{+5.33} & \ns{+0.38} & \ns{+1.43}
& \sig{+4.07} & \ns{+0.38} & \ns{+0.48}
& \ns{+0.00} & \ns{+0.57} & \ns{+0.76} \\

\addlinespace[2pt]

\multirow{4}{*}{Style}
& Fluency
& \ns{+0.26} & \ns{+0.87} & \ns{$-$0.35}
& \ns{+1.07} & \ns{$-$0.19} & \sig{+2.57}
& \ns{+1.90} & \ns{$-$1.71} & \ns{+0.29}
& \ns{$-$0.81} & \ns{+0.57} & \ns{+0.19} \\

& Simple Language
& \ns{+0.13} & \ns{+0.79} & \ns{$-$0.09}
& \ns{+2.93} & \ns{$-$0.95} & \ns{+1.14}
& \ns{+1.35} & \sig{$-$3.62} & \ns{$-$0.86}
& \ns{+1.08} & \ns{+0.95} & \ns{+0.95} \\

& Technical Terms
& \ns{$-$0.13} & \ns{+0.26} & \sig{$-$1.66}
& \ns{+1.07} & \ns{+0.19} & \sig{+2.38}
& \ns{$-$0.27} & \ns{$-$2.10} & \ns{$-$0.38}
& \ns{+0.00} & \ns{+0.76} & \ns{$-$0.57} \\

& Unique Words
& \ns{+0.22} & \ns{+0.57} & \ns{+0.04}
& \ns{+1.33} & \ns{$-$0.57} & \ns{+1.81}
& \ns{+1.63} & \ns{+0.19} & \ns{+1.05}
& \ns{$-$0.54} & \ns{+0.95} & \ns{$-$0.57} \\

\midrule
\multicolumn{14}{l}{\textit{Share of Voice (SOV)} ($\Delta$, pp)} \\
\midrule

& \textit{Baseline (\%)}
& 56.7 & 53.1 & 53.5
& 16.1 & 10.9 & 11.2
& 7.2 & 12.0 & 10.2
& 1.6 & 3.4 & 4.6 \\

& \textit{Placebo}
& \ns{$-$0.33} & \ns{$-$0.36} & \ns{+0.34}
& \ns{$-$0.06} & \ns{+0.31} & \ns{$-$0.17}
& \ns{$-$0.34} & \ns{$-$0.63} & \ns{+0.15}
& \ns{$-$0.01} & \ns{+0.16} & \ns{+0.09} \\

\addlinespace[2pt]

\multirow{2}{*}{Argumentation}
& Comparative Distinction
& \sig{+1.34} & \sig{+1.07} & \sig{+1.18}
& \sig{+5.15} & \ns{+0.06} & \sig{+0.69}
& \sig{+2.94} & \ns{+0.60} & \sig{+0.57}
& \sig{+1.69} & \ns{$-$0.03} & \sig{+0.41} \\

& Refutational
& \sig{+1.02} & \ns{+0.39} & \sig{+2.77}
& \sig{+2.08} & \ns{+0.58} & \ns{+0.27}
& \ns{+0.27} & \ns{$-$0.19} & \sig{+0.65}
& \ns{+0.46} & \ns{+0.44} & \sig{+0.33} \\

\addlinespace[2pt]

\multirow{2}{*}{Structure}
& Content Improvement
& \sig{+0.81} & \sig{+1.33} & \sig{+3.23}
& \sig{+3.08} & \ns{+0.67} & \ns{+0.33}
& \sig{+1.96} & \ns{+0.14} & \sig{+0.37}
& \ns{+0.78} & \sig{+0.52} & \sig{+0.53} \\

& LLMs.txt
& \sig{+0.80} & \ns{+0.78} & \sig{+2.32}
& \sig{+3.53} & \sig{+1.51} & \sig{+0.47}
& \sig{+2.75} & \ns{+0.40} & \sig{+0.47}
& \sig{+1.81} & \sig{+0.76} & \sig{+1.00} \\

\addlinespace[2pt]

\multirow{3}{*}{Evidence}
& Quotes
& \sig{+0.78} & \ns{+0.45} & \ns{+0.46}
& \sig{+3.12} & \ns{$-$0.25} & \sig{+0.52}
& \sig{+2.28} & \ns{$-$0.10} & \ns{+0.11}
& \ns{+0.55} & \ns{+0.45} & \ns{+0.25} \\

& Statistics
& \sig{+1.00} & \sig{+1.17} & \sig{+3.71}
& \ns{+1.48} & \ns{$-$0.15} & \ns{$-$0.16}
& \ns{+0.64} & \ns{+0.31} & \ns{$-$0.12}
& \ns{+0.18} & \ns{+0.16} & \ns{+0.28} \\

& Authoritative
& \ns{+0.03} & \ns{+0.53} & \sig{+1.68}
& \ns{+0.49} & \ns{$-$0.09} & \ns{+0.02}
& \sig{+1.06} & \ns{$-$0.13} & \ns{+0.14}
& \ns{$-$0.09} & \ns{+0.19} & \ns{$-$0.07} \\

\addlinespace[2pt]

\multirow{4}{*}{Style}
& Fluency
& \ns{$-$0.13} & \ns{+0.74} & \sig{+2.80}
& \ns{+0.46} & \ns{+0.03} & \ns{$-$0.03}
& \ns{+0.69} & \sig{$-$0.82} & \ns{+0.03}
& \ns{$-$0.56} & \ns{+0.02} & \ns{+0.16} \\

& Simple Language
& \ns{$-$0.01} & \ns{+0.07} & \sig{+2.40}
& \sig{+1.69} & \ns{+0.17} & \ns{$-$0.16}
& \ns{+0.23} & \ns{$-$0.34} & \ns{+0.19}
& \ns{+0.11} & \ns{$-$0.16} & \ns{+0.21} \\

& Technical Terms
& \ns{$-$0.27} & \ns{+0.58} & \sig{+2.10}
& \ns{$-$0.64} & \ns{+0.02} & \ns{+0.04}
& \ns{+0.10} & \ns{$-$0.17} & \ns{+0.18}
& \ns{$-$0.16} & \ns{+0.03} & \ns{$-$0.08} \\

& Unique Words
& \ns{+0.16} & \ns{+0.11} & \sig{+1.41}
& \ns{+0.66} & \ns{+0.14} & \ns{$-$0.14}
& \ns{+0.14} & \ns{$-$0.34} & \ns{+0.09}
& \ns{$-$0.20} & \ns{+0.03} & \ns{+0.02} \\

\midrule
\multicolumn{14}{l}{\textit{Win Rate} ($\Delta$, pp)} \\
\midrule

& \textit{Baseline (\%)}
& 63.6 & 54.4 & 58.6
& 16.3 & 12.9 & 13.2
& 7.1 & 11.5 & 8.9
& 0.9 & 3.0 & 2.5 \\

& \textit{Placebo}
& \sig{$-$1.38} & \ns{$-$0.68} & \ns{+0.33}
& \ns{+0.07} & \ns{$-$0.92} & \ns{$-$0.18}
& \ns{$-$1.58} & \sig{$-$2.11} & \ns{+0.02}
& \ns{$-$0.41} & \ns{$-$0.41} & \ns{+0.38} \\

\addlinespace[2pt]

\multirow{2}{*}{Argumentation}
& Comparative Distinction
& \sig{+3.08} & \sig{+1.46} & \sig{+2.53}
& \sig{+9.80} & \ns{$-$0.55} & \ns{+1.81}
& \ns{+2.78} & \ns{+0.49} & \ns{+1.16}
& \ns{+1.13} & \ns{$-$0.41} & \sig{+1.46} \\

& Refutational
& \sig{+1.66} & \ns{+0.87} & \sig{+2.21}
& \sig{+4.27} & \ns{$-$1.24} & \ns{+0.45}
& \ns{$-$0.32} & \ns{$-$0.90} & \ns{+0.92}
& \ns{+0.14} & \ns{+0.02} & \ns{+0.20} \\

\addlinespace[2pt]

\multirow{2}{*}{Structure}
& Content Improvement
& \sig{+1.90} & \ns{+1.07} & \sig{+3.71}
& \sig{+4.48} & \ns{+1.28} & \ns{+0.46}
& \ns{+2.08} & \ns{$-$0.15} & \ns{+1.39}
& \ns{+1.15} & \ns{$-$0.49} & \sig{+1.58} \\

& LLMs.txt
& \sig{+2.18} & \ns{+0.74} & \sig{+2.05}
& \sig{+5.07} & \ns{+1.81} & \ns{+0.10}
& \ns{+1.26} & \ns{$-$0.32} & \ns{+0.60}
& \ns{$-$0.07} & \ns{+0.69} & \sig{+2.38} \\

\addlinespace[2pt]

\multirow{3}{*}{Evidence}
& Quotes
& \ns{+1.11} & \ns{+0.15} & \ns{+1.09}
& \sig{+5.92} & \ns{$-$2.13} & \ns{+0.79}
& \ns{+1.26} & \ns{$-$0.42} & \ns{$-$0.35}
& \ns{$-$0.27} & \ns{$-$0.52} & \ns{+0.94} \\

& Statistics
& \sig{+1.90} & \ns{+0.98} & \sig{+4.46}
& \ns{+2.02} & \ns{$-$0.37} & \ns{$-$0.15}
& \ns{$-$0.90} & \ns{+0.29} & \ns{$-$0.24}
& \ns{$-$0.02} & \ns{$-$0.86} & \ns{+1.06} \\

& Authoritative
& \ns{$-$0.24} & \ns{+0.83} & \ns{+0.63}
& \ns{+0.73} & \ns{$-$0.69} & \ns{$-$1.04}
& \ns{$-$0.93} & \ns{$-$0.11} & \ns{$-$0.49}
& \ns{$-$0.14} & \ns{$-$0.73} & \ns{+0.02} \\

\addlinespace[2pt]

\multirow{4}{*}{Style}
& Fluency
& \ns{$-$0.35} & \ns{+0.79} & \ns{+1.55}
& \ns{+2.22} & \ns{$-$0.83} & \ns{$-$0.35}
& \ns{$-$0.27} & \ns{$-$1.12} & \ns{$-$0.01}
& \ns{$-$0.81} & \ns{$-$0.71} & \ns{+0.70} \\

& Simple Language
& \ns{+0.20} & \ns{$-$0.07} & \ns{+1.25}
& \sig{+4.27} & \ns{$-$1.05} & \ns{$-$1.03}
& \ns{$-$1.17} & \ns{$-$0.32} & \ns{$-$0.65}
& \ns{$-$0.54} & \ns{$-$0.87} & \ns{+0.93} \\

& Technical Terms
& \ns{$-$0.72} & \ns{+0.37} & \ns{+1.51}
& \ns{$-$1.82} & \ns{$-$2.00} & \ns{$-$0.87}
& \ns{$-$0.59} & \ns{$-$0.47} & \ns{$-$0.29}
& \ns{$-$0.41} & \ns{$-$0.86} & \ns{+0.66} \\

& Unique Words
& \ns{$-$0.72} & \ns{+0.55} & \ns{+1.07}
& \ns{+2.42} & \ns{$-$0.62} & \ns{$-$1.44}
& \ns{$-$1.13} & \ns{$-$2.03} & \ns{+0.47}
& \ns{$-$0.14} & \sig{$-$1.21} & \ns{+0.85} \\

\bottomrule
\end{tabular}
\end{table*}

\begin{table*}[!t]
\centering
\caption{Full persona results in the U.S. across response engines and the two strongest rewriting strategies. Base reports exposure before rewriting under each persona, and all other entries report paired changes from that baseline in percentage points. Black values are statistically significant ($p<0.05$), while gray values are not. Placebo compares two generations without rewriting. CI = Content Improvement; CD = Comparative Distinction. Persona-free denotes generation without a persona.}
\label{tab:full-persona-strategy-engine}
\vspace{-2mm}

\footnotesize
\setlength{\tabcolsep}{2.4pt}
\renewcommand{\arraystretch}{1.08}

\begin{tabular}{llrrrrrrrrrrrr}
\toprule
& &
\multicolumn{4}{c}{\textbf{\gptfourmini}} &
\multicolumn{4}{c}{\textbf{\gptfivemini}} &
\multicolumn{4}{c}{\textbf{\gemini}} \\
\cmidrule(lr){3-6}
\cmidrule(lr){7-10}
\cmidrule(lr){11-14}

\textbf{Category}
& \textbf{Persona}
& \textbf{Base} & \textbf{Placebo} & \textbf{CI} & \textbf{CD}
& \textbf{Base} & \textbf{Placebo} & \textbf{CI} & \textbf{CD}
& \textbf{Base} & \textbf{Placebo} & \textbf{CI} & \textbf{CD} \\
\midrule

\multicolumn{14}{l}{\textit{Mention Coverage} (Base in \%, others $\Delta$ in pp)} \\
\midrule

-- & \textit{Persona-free}
& 90.7 & \ns{$-$0.26} & \ns{+0.57} & \ns{+0.31}
& 67.9 & \ns{+0.74} & \sig{+2.14} & \sig{+2.62}
& 89.2 & \ns{+0.00} & \ns{$-$0.31} & \ns{+0.44} \\

\addlinespace[2pt]
\multirow{9}{*}{Demographic identity}
& White
& 89.9 & \ns{+0.09} & \ns{+0.70} & \sig{+1.40}
& 70.9 & \ns{$-$0.39} & \sig{+2.05} & \ns{+1.44}
& 89.6 & \ns{+0.04} & \ns{+0.79} & \ns{+0.96} \\

& Black
& 90.3 & \ns{+0.52} & \sig{+1.27} & \sig{+1.27}
& 72.8 & \ns{$-$0.79} & \sig{+2.14} & \ns{+0.17}
& 91.2 & \ns{$-$0.61} & \ns{$-$0.31} & \ns{+0.09} \\

& Hispanic
& 90.2 & \ns{+0.04} & \ns{+0.87} & \sig{+1.49}
& 72.1 & \ns{+0.09} & \ns{+0.96} & \ns{+1.18}
& 90.3 & \ns{$-$0.22} & \ns{$-$0.61} & \ns{+0.39} \\

& Asian
& 90.0 & \ns{+0.04} & \sig{+1.18} & \ns{+1.05}
& 72.1 & \ns{$-$0.04} & \ns{+1.22} & \ns{+0.83}
& 90.1 & \ns{$-$0.04} & \ns{+1.27} & \sig{+1.62} \\

& American Indian
& 89.9 & \ns{+0.17} & \ns{+0.61} & \ns{+0.74}
& 73.0 & \ns{$-$0.92} & \ns{+1.00} & \ns{$-$0.57}
& 90.3 & \ns{+0.35} & \ns{$-$0.04} & \ns{+0.22} \\

& Pacific Islander
& 90.2 & \ns{+0.35} & \ns{+0.48} & \ns{+0.70}
& 72.2 & \ns{+0.87} & \sig{+1.75} & \sig{+1.97}
& 89.7 & \ns{$-$0.66} & \ns{+0.74} & \ns{+0.92} \\

& Man
& 90.5 & \ns{+0.39} & \ns{+0.44} & \ns{+0.35}
& 71.1 & \ns{$-$0.83} & \sig{+1.62} & \sig{+2.49}
& 88.9 & \ns{$-$0.44} & \ns{+0.79} & \ns{+0.96} \\

& Woman
& 90.6 & \ns{+0.04} & \sig{+1.05} & \ns{+0.83}
& 72.9 & \sig{$-$2.05} & \ns{+0.74} & \ns{$-$0.04}
& 90.6 & \ns{$-$0.52} & \ns{$-$0.13} & \ns{+0.35} \\

& Non-binary
& 90.1 & \ns{+0.48} & \sig{+1.31} & \sig{+1.09}
& 72.0 & \sig{$-$1.83} & \ns{$-$0.61} & \ns{$-$0.22}
& 90.0 & \ns{+0.22} & \ns{+1.00} & \ns{+0.26} \\

\addlinespace[2pt]
\multirow{3}{*}{Partisan identity}
& Democrat
& 91.7 & \ns{+0.61} & \ns{+0.83} & \sig{+1.27}
& 75.5 & \ns{$-$0.26} & \ns{+0.31} & \ns{+1.27}
& 90.6 & \ns{+0.00} & \ns{+0.87} & \sig{+1.57} \\

& Independent
& 90.4 & \ns{$-$0.66} & \ns{+0.48} & \ns{+0.79}
& 74.5 & \ns{$-$0.87} & \ns{+1.00} & \ns{+0.44}
& 90.7 & \ns{+0.13} & \ns{$-$0.04} & \ns{+0.57} \\

& Republican
& 88.4 & \ns{$-$0.22} & \ns{+0.52} & \ns{+0.74}
& 71.8 & \ns{$-$1.31} & \ns{+1.35} & \ns{+0.79}
& 90.3 & \ns{$-$0.04} & \ns{$-$0.31} & \ns{$-$0.26} \\

\addlinespace[2pt]
\multirow{3}{*}{Questionnaire-derived profile}
& Lean left
& 89.6 & \ns{+0.04} & \sig{+1.31} & \ns{+0.57}
& 74.7 & \ns{+0.31} & \sig{+1.75} & \ns{+1.44}
& 92.4 & \ns{$-$0.39} & \ns{$-$0.92} & \ns{$-$0.74} \\

& Swing
& 89.3 & \ns{$-$0.48} & \ns{+0.09} & \ns{+0.26}
& 74.5 & \ns{$-$1.22} & \ns{+0.87} & \ns{+0.26}
& 90.2 & \ns{$-$0.17} & \ns{+0.96} & \ns{+0.70} \\

& Lean right
& 89.3 & \ns{$-$0.04} & \sig{+1.05} & \sig{+1.27}
& 73.5 & \ns{$-$0.13} & \ns{+1.49} & \ns{+1.22}
& 89.5 & \ns{$-$0.44} & \ns{+0.96} & \ns{+0.96} \\

\midrule
\multicolumn{14}{l}{\textit{Share of Voice (SOV)} (Base in \%, others $\Delta$ in pp)} \\
\midrule

-- & \textit{Persona-free}
& 56.5 & \ns{$-$0.07} & \sig{+0.70} & \sig{+1.42}
& 50.1 & \ns{+0.44} & \sig{+1.14} & \sig{+1.81}
& 56.0 & \ns{$-$0.02} & \ns{+0.30} & \sig{+1.53} \\

\addlinespace[2pt]
\multirow{9}{*}{Demographic identity}
& White
& 55.7 & \sig{+0.56} & \sig{+1.54} & \sig{+2.45}
& 51.5 & \ns{$-$0.37} & \ns{+0.89} & \ns{+0.76}
& 56.3 & \ns{+0.30} & \ns{+0.57} & \sig{+1.41} \\

& Black
& 58.0 & \ns{+0.30} & \ns{+0.56} & \sig{+2.04}
& 52.2 & \ns{$-$0.54} & \sig{+1.38} & \ns{+0.59}
& 57.2 & \ns{$-$0.29} & \ns{+0.59} & \sig{+1.78} \\

& Hispanic
& 57.6 & \ns{$-$0.21} & \sig{+0.65} & \sig{+1.77}
& 51.8 & \ns{+0.27} & \sig{+1.41} & \sig{+1.50}
& 56.3 & \ns{$-$0.06} & \sig{+1.02} & \sig{+2.25} \\

& Asian
& 57.2 & \ns{$-$0.23} & \sig{+0.91} & \sig{+2.09}
& 51.9 & \ns{+0.24} & \sig{+1.19} & \ns{+0.53}
& 56.5 & \ns{+0.12} & \sig{+1.30} & \sig{+1.89} \\

& American Indian
& 56.9 & \ns{+0.24} & \sig{+1.09} & \sig{+1.90}
& 52.6 & \ns{$-$0.49} & \ns{+0.46} & \ns{+0.16}
& 56.5 & \ns{+0.27} & \ns{+0.64} & \sig{+1.58} \\

& Pacific Islander
& 57.1 & \ns{+0.20} & \sig{+0.71} & \sig{+1.74}
& 52.3 & \ns{+0.50} & \sig{+1.30} & \sig{+1.14}
& 56.4 & \ns{$-$0.20} & \sig{+1.26} & \sig{+2.10} \\

& Man
& 56.1 & \ns{+0.36} & \sig{+0.95} & \sig{+1.44}
& 51.4 & \ns{$-$0.23} & \sig{+1.01} & \sig{+1.64}
& 56.2 & \sig{$-$0.40} & \sig{+0.79} & \sig{+1.61} \\

& Woman
& 57.4 & \ns{+0.34} & \sig{+1.02} & \sig{+2.10}
& 52.0 & \ns{$-$0.61} & \sig{+1.15} & \sig{+1.14}
& 56.5 & \ns{$-$0.10} & \ns{+0.68} & \sig{+1.72} \\

& Non-binary
& 56.8 & \ns{+0.08} & \sig{+0.89} & \sig{+1.74}
& 51.7 & \ns{$-$0.55} & \ns{$-$0.22} & \ns{+0.58}
& 56.5 & \ns{$-$0.09} & \sig{+0.90} & \sig{+1.33} \\

\addlinespace[2pt]
\multirow{3}{*}{Partisan identity}
& Democrat
& 61.1 & \ns{+0.22} & \ns{+0.41} & \sig{+2.20}
& 54.3 & \ns{+0.38} & \sig{+1.08} & \sig{+1.76}
& 57.8 & \ns{+0.07} & \sig{+0.76} & \sig{+2.16} \\

& Independent
& 56.8 & \ns{$-$0.32} & \ns{+0.49} & \sig{+1.57}
& 53.3 & \ns{$-$0.40} & \ns{+0.80} & \ns{+0.69}
& 56.9 & \ns{$-$0.26} & \ns{+0.30} & \sig{+1.64} \\

& Republican
& 49.5 & \ns{+0.28} & \ns{+0.27} & \sig{+1.36}
& 50.9 & \ns{$-$0.38} & \sig{+0.95} & \ns{+0.78}
& 54.8 & \ns{$-$0.14} & \sig{+0.84} & \sig{+1.21} \\

\addlinespace[2pt]
\multirow{3}{*}{Questionnaire-derived profile}
& Lean left
& 56.9 & \ns{+0.03} & \ns{+0.45} & \sig{+1.45}
& 54.2 & \ns{$-$0.04} & \sig{+0.95} & \sig{+1.11}
& 57.1 & \ns{$-$0.21} & \ns{+0.22} & \sig{+1.15} \\

& Swing
& 57.1 & \ns{$-$0.09} & \ns{$-$0.05} & \sig{+0.86}
& 54.1 & \ns{$-$0.37} & \ns{+0.91} & \ns{+0.72}
& 56.8 & \ns{+0.11} & \sig{+0.77} & \sig{+1.25} \\

& Lean right
& 56.9 & \ns{$-$0.05} & \sig{+0.80} & \sig{+1.50}
& 53.4 & \ns{+0.20} & \sig{+0.97} & \sig{+1.36}
& 56.1 & \ns{$-$0.14} & \sig{+0.92} & \sig{+1.82} \\

\midrule
\multicolumn{14}{l}{\textit{Win Rate} (Base in \%, others $\Delta$ in pp)} \\
\midrule

-- & \textit{Persona-free}
& 63.1 & \ns{$-$0.13} & \sig{+1.97} & \sig{+2.71}
& 51.1 & \ns{+0.52} & \sig{+1.38} & \sig{+2.34}
& 59.5 & \ns{+0.02} & \ns{+0.50} & \sig{+3.30} \\

\addlinespace[2pt]
\multirow{9}{*}{Demographic identity}
& White
& 63.1 & \ns{$-$0.48} & \sig{+2.62} & \sig{+3.23}
& 52.3 & \ns{$-$0.63} & \ns{+1.18} & \ns{+0.92}
& 60.4 & \ns{$-$0.63} & \ns{$-$0.04} & \sig{+1.83} \\

& Black
& 66.7 & \ns{+0.59} & \ns{+0.94} & \sig{+3.82}
& 53.0 & \ns{$-$0.35} & \sig{+1.94} & \ns{+1.18}
& 61.1 & \ns{$-$0.37} & \ns{+1.11} & \sig{+3.74} \\

& Hispanic
& 66.0 & \ns{$-$0.70} & \ns{+0.76} & \sig{+3.58}
& 52.8 & \ns{+0.20} & \sig{+1.51} & \sig{+1.77}
& 59.9 & \ns{$-$0.26} & \ns{+1.33} & \sig{+3.54} \\

& Asian
& 65.2 & \ns{$-$0.96} & \ns{+1.07} & \sig{+3.89}
& 52.6 & \ns{+0.28} & \sig{+2.18} & \ns{+1.03}
& 59.5 & \ns{$-$0.11} & \ns{+1.40} & \sig{+3.78} \\

& American Indian
& 63.7 & \ns{+0.42} & \sig{+2.10} & \sig{+4.83}
& 53.8 & \ns{$-$0.28} & \ns{+0.55} & \ns{+0.63}
& 60.0 & \ns{+0.28} & \ns{+1.09} & \sig{+3.52} \\

& Pacific Islander
& 63.4 & \ns{+1.16} & \sig{+2.71} & \sig{+4.48}
& 53.5 & \ns{+0.26} & \sig{+1.33} & \sig{+1.55}
& 60.5 & \ns{$-$0.37} & \ns{+1.38} & \sig{+3.49} \\

& Man
& 61.3 & \sig{+1.92} & \sig{+2.88} & \sig{+5.09}
& 52.1 & \ns{$-$0.15} & \ns{+0.94} & \sig{+1.53}
& 58.7 & \ns{$-$0.33} & \sig{+2.80} & \sig{+4.43} \\

& Woman
& 64.3 & \sig{+1.62} & \sig{+2.93} & \sig{+4.02}
& 52.5 & \ns{$-$0.98} & \sig{+1.57} & \sig{+1.57}
& 59.1 & \ns{$-$0.24} & \sig{+2.84} & \sig{+4.91} \\

& Non-binary
& 63.5 & \ns{$-$0.57} & \sig{+1.97} & \sig{+4.30}
& 52.0 & \ns{$-$0.52} & \ns{+0.61} & \sig{+2.01}
& 60.0 & \ns{$-$0.57} & \sig{+2.18} & \sig{+3.32} \\

\addlinespace[2pt]
\multirow{3}{*}{Partisan identity}
& Democrat
& 71.4 & \ns{+0.02} & \ns{+0.50} & \sig{+2.97}
& 56.0 & \ns{$-$0.02} & \ns{+0.90} & \sig{+1.92}
& 62.2 & \ns{$-$0.22} & \sig{+1.97} & \sig{+4.22} \\

& Independent
& 63.0 & \ns{$-$0.09} & \ns{+0.81} & \sig{+2.97}
& 54.3 & \ns{$-$0.37} & \ns{+0.39} & \ns{+1.22}
& 60.9 & \sig{$-$0.96} & \ns{+0.37} & \sig{+2.21} \\

& Republican
& 50.8 & \ns{+0.31} & \ns{$-$0.83} & \ns{+1.11}
& 51.7 & \ns{$-$0.83} & \ns{+0.72} & \ns{+0.79}
& 56.0 & \ns{+0.26} & \sig{+1.86} & \sig{+3.39} \\

\addlinespace[2pt]
\multirow{3}{*}{Questionnaire-derived profile}
& Lean left
& 64.5 & \ns{$-$0.72} & \ns{+0.76} & \ns{+1.57}
& 54.7 & \ns{+0.52} & \sig{+1.86} & \sig{+1.81}
& 61.0 & \ns{$-$0.35} & \ns{+0.50} & \sig{+2.23} \\

& Swing
& 64.5 & \ns{+0.13} & \ns{+0.28} & \sig{+2.77}
& 55.0 & \ns{$-$0.83} & \ns{+0.92} & \ns{+0.35}
& 60.6 & \ns{+0.46} & \ns{+0.37} & \ns{+1.62} \\

& Lean right
& 63.9 & \ns{$-$1.27} & \sig{+1.62} & \sig{+3.17}
& 54.4 & \ns{$-$0.31} & \ns{+0.83} & \sig{+1.83}
& 58.0 & \ns{$-$0.13} & \sig{+2.47} & \sig{+4.70} \\

\bottomrule
\end{tabular}
\end{table*}

\subsection{Inference Robustness}
\label{sec:app-inference}

Because multiple daily cells share the same base query, we repeat the main tests using standard errors clustered by base query, with 109 clusters for the U.S. and 25 for Japan, and apply Benjamini--Hochberg correction at a 5\% false discovery rate within each table.
For RQ1, 18 of 36 target-specified comparisons remain significant after clustering and correction, compared with 24 originally, while none of the 36 target-unspecified comparisons remain significant, compared with 5 originally. For RQ2, 22 of 48 category--election-group--metric comparisons remain significant, compared with 31 originally. The nine lost comparisons have absolute effects below 1.5 percentage points, while the SOV ordering Structure $\geq$ Argumentation $>$ Evidence $>$ Style remains unchanged in the U.S. and all three Japanese tiers.
For RQ3, none of the 90 persona $\mathrm{Mod}$ contrasts remain significant after clustering and correction, compared with 7 originally and 5 with clustering alone. The originally significant contrasts also have mixed directions across engines. Thus, persona conditioning shows no consistent tendency to increase or decrease rewriting gains. These checks leave the qualitative conclusions unchanged.

\subsection{Full Results}
\label{sec:full-results}

Tables~\ref{tab:full-strategy-election-engine} and~\ref{tab:full-persona-strategy-engine} report the full effect sizes underlying Section~\ref{sec:exp}, including baseline and placebo references. Table~\ref{tab:rq3-jp-gender} further reports gender-persona results across three Japanese party tiers using \texttt{gpt-4o-mini}. Clustered and multiple-comparison robustness is reported in Section~\ref{sec:app-inference}.

\end{document}